\documentclass[aps,prd,twocolumn,showpacs,amsmath,amssymb]{revtex4-1}
\usepackage{amsmath}
\usepackage{graphicx}
\usepackage{subfigure}
\usepackage{epstopdf}
\usepackage{color}
\usepackage{multirow}
\usepackage{setspace}
\usepackage{overpic}
\usepackage{amssymb}
\usepackage[driverfallback=dvipdfmx, bookmarksnumbered, pdfstartview=FitH,colorlinks,urlcolor=blue, citecolor=blue,linkcolor=blue] {hyperref}
\usepackage{lineno}
\usepackage{bm}
\usepackage{makecell}
\usepackage{rotating}
\usepackage[utf8]{inputenc}

\let\oldequation\equation
\let\oldendequation\endequation

\renewenvironment{equation}
  {\linenomathNonumbers\oldequation}
  {\oldendequation\endlinenomath}

\newcommand{\tpi}{\pi^+\pi^-\pi^0}

\newcommand{\etaev}{D_s^+\to \eta e^+\nu_e}
\newcommand{\etapev}{D_s^+\to \eta^\prime e^+\nu_e}
\newcommand{\phiev}{D_s^+\to \phi e^+\nu_e}
\newcommand{\foev}{D_s^+\to f_0 e^+\nu_e}
\newcommand{\Koev}{D_s^+\to K^0 e^+\nu_e}
\newcommand{\Kstoev}{D_s^+\to K^{*0} e^+\nu_e}

\newcommand{\bfetaev}{2.35\pm0.11_{\rm stat}\pm0.10_{\rm syst}}
\newcommand{\bfetapev}{0.82\pm0.09_{\rm stat}\pm0.04_{\rm syst}}

\newcommand{\bfphiev}{2.21\pm0.16_{\rm stat}\pm0.11_{\rm syst}}
\newcommand{\bffopev}{0.15\pm0.02_{\rm stat}\pm0.01_{\rm syst}}
\newcommand{\bfKoev}{0.24\pm0.04_{\rm stat}\pm0.01_{\rm syst}}
\newcommand{\bfKstoev}{0.19\pm0.03_{\rm stat}\pm0.01_{\rm syst}}

\newcommand{\czetaevnew}{1.7/3}

\newcommand{\czetapevnew}{1.0/1}

\newcommand{\czkoevnew}{0.1/1}

\newcommand{\fvzetaevnewnew}{0.430\pm0.021\pm0.016}
\newcommand{\rzetaevnewnew}{-4.7\pm1.0\pm0.2}

\newcommand{\fzetaevnewnew}{0.442\pm0.022\pm0.017}

\newcommand{\fvzetapevnewnew}{0.542\pm0.062\pm0.023}
\newcommand{\rzetapevnewnew}{-4.0\pm9.5\pm1.7}

\newcommand{\fzetapevnewnew}{0.557\pm0.062\pm0.024}

\newcommand{\fvzkoevnewnew}{0.152\pm0.022\pm0.005}
\newcommand{\rzkoevnewnew}{-0.1\pm3.4\pm0.6}

\newcommand{\fzkoevnewnew}{0.677\pm0.098\pm0.023}

\begin{document}

\title{\bf \boldmath
Measurements of the branching fractions of semileptonic $D^{+}_s$ decays via $e^+e^-\to D_s^{*+}D_s^{*-}$
}

\author{
M.~Ablikim$^{1}$, M.~N.~Achasov$^{4,c}$, P.~Adlarson$^{76}$, O.~Afedulidis$^{3}$, X.~C.~Ai$^{81}$, R.~Aliberti$^{35}$, A.~Amoroso$^{75A,75C}$, Q.~An$^{72,58,a}$, Y.~Bai$^{57}$, O.~Bakina$^{36}$
% ...
	\\
	\vspace{0.2cm}
	(BESIII Collaboration)\\
	\vspace{0.2cm} {\it
		$^{1}$ Institute of High Energy Physics, Beijing 100049, People's Republic of China\\
% ...
$^{3}$ Bochum Ruhr-University, D-44780 Bochum, Germany\\
		$^{4}$ Budker Institute of Nuclear Physics SB RAS (BINP), Novosibirsk 630090, Russia\\
% ...
$^{35}$ Johannes Gutenberg University of Mainz, Johann-Joachim-Becher-Weg 45, D-55099 Mainz, Germany\\
		$^{36}$ Joint Institute for Nuclear Research, 141980 Dubna, Moscow region, Russia\\
% ...
$^{57}$ Southeast University, Nanjing 211100, People's Republic of China\\
		$^{58}$ State Key Laboratory of Particle Detection and Electronics, Beijing 100049, Hefei 230026, People's Republic of China\\
% ...
$^{72}$ University of Science and Technology of China, Hefei 230026, People's Republic of China\\
% ...
$^{75}$ University of Turin and INFN, (A)University of Turin, I-10125, Turin, Italy; (B)University of Eastern Piedmont, I-15121, Alessandria, Italy; (C)INFN, I-10125, Turin, Italy\\
		$^{76}$ Uppsala University, Box 516, SE-75120 Uppsala, Sweden\\
% ...
$^{81}$ Zhengzhou University, Zhengzhou 450001, People's Republic of China\\
		\vspace{0.2cm}
		$^{a}$ Deceased\\
% ...
$^{c}$ Also at the Novosibirsk State University, Novosibirsk, 630090, Russia\\
% ...
		}
}

\begin{abstract}
We measure the absolute branching fractions of semileptonic $D^+_s$
decays via the $e^+e^-\to D_s^{*+}D_s^{*-}$ process using $e^+e^-$
collision data corresponding to an integrated luminosity of
$10.64~\mathrm{fb}^{-1}$ collected by the BESIII detector at
center-of-mass energies between 4.237 and 4.699 GeV.  The branching
fractions are ${\mathcal B}(\etaev)=(\bfetaev)\%,$ ${\mathcal
	B}(\etapev)=(\bfetapev)\%,$ ${\mathcal B}(\phiev)=(\bfphiev)\%,$ 
${\mathcal B}(D_s^+\to f_0(980)
e^+\nu_e,f_0(980)\to\pi^+\pi^-)=(\bffopev)\%,$ ${\mathcal
	B}(\Koev)=(\bfKoev)\%,$ and ${\mathcal B}(\Kstoev)=(\bfKstoev)\%.$
These results are consistent with those measured via the $e^+e^-\to
D_s^{*\pm}D_s^{\mp}$ process by BESIII and CLEO. Using two-parameter series expansion, the hadronic transition form factors of $D^+_s\to \eta e^+\nu_e$, $D^+_s\to \eta^\prime
e^+\nu_e$, and $D^+_s\to K^0 e^+\nu_e$ are determined to be $f^{\eta}_+(0) = 0.442\pm 0.022_{\rm stat}\pm 0.017_{\rm syst},$
$f^{\eta^{\prime}}_+(0) = 0.557\pm 0.062_{\rm stat}\pm0.024_{\rm syst},$ and 
$f^{K^0}_+(0) = 0.677\pm0.098_{\rm stat}\pm0.023_{\rm syst}.$
\end{abstract}

\maketitle

\oddsidemargin  -0.2cm
\evensidemargin -0.2cm

\section{Introduction}

	Experimental studies of semileptonic $D^+_s$ decays are important to
	understand the weak and strong effects in charm quark decays. 
	By analyzing their decay dynamics, one
	can extract the product of the modulus of the Cabibbo-Kobayashi-Maskawa (CKM) matrix element $|V_{cs(d)}|$ and
	the hadronic transition form factor, providing valuable insights
	into charm physics. 
	Studies of these decays offer opportunity to
determine  hadronic transition form factors by inputting   the $|V_{cs(d)}|$ from the standard model global fit. 
The  hadronic
form factors obtained are valuable to test  theoretical calculations.  Moreover, different
frameworks~\cite{Melikhov:2000yu,RQM:2020,chisq:2005,Verma:2011yw,Wei:2009nc,Offen:2013nma,Soni:2018adu,Hu:2021zmy,H:2017,X:2021,Y:2006,Soni2020,LCSR2,LCSR1},
e.g., quark model, QCD sum rule, and lattice QCD, provide predictions
on the branching fractions. Table~\ref{tab:theoryBF} summarizes the
branching fractions of semileptonic $D^+_s$ decays predicted by
various theoretical models
~\cite{Melikhov:2000yu,RQM:2020,chisq:2005,Verma:2011yw,Wei:2009nc,Offen:2013nma,Soni:2018adu,Hu:2021zmy,H:2017,X:2021,Y:2006,Soni2020,LCSR2,LCSR1}. Precise
measurements of these decay branching fractions are useful to provide
tighter constraints on theory.

Since 2008, the CLEO~\cite{cleo:4170} and BESIII
Collaborations~\cite{Ke:2023qzc} have reported measurements of the
branching fractions of the semileptonic $D_s^+$ decays, as summarized
in the Particle Data Group (PDG)~\cite{pdg2024}.  These measurements
are performed by using the $e^+e^-\to D^+_sD^-_s$ and $e^+e^-\to
D^{*\pm}_sD^{\mp}_s$ processes with 0.48 fb$^{-1}$ and 7.33 fb$^{-1}$
of $e^+e^-$ collision data taken at center-of-mass energies of $\sqrt
s=4.009$ and $4.128{\text -}4.226$ GeV, respectively.  In this paper,
we report the measurements of the branching fractions of the
semileptonic $D^+_s$ decays via the $e^+e^-\to D_s^{*+}D_s^{*-}$
process, based on the analysis of 10.64 fb$^{-1}$ of $e^+e^-$
collision data taken at $\sqrt s=4.237{\text -}4.699$ GeV with the
BESIII detector. Throughout this paper, charge conjugation is always
implied, and $\rho$, $K^{*0}$, and $f_0$ denote the $\rho(770)$,
$K^*(892)^0$, and $f_0(980)$, respectively.

\begin{table*}[htbp]
	\centering
	\caption{The branching fractions (in percent) of the semileptonic $D^+_s$ decays predicted by various theories.}
	\label{tab:theoryBF}
	\scalebox{1.0}{
		\begin{tabular}{l|cccccc}
			\hline
			\hline
				& $\etaev$ & $\etapev$  & $\phiev$ & $D_s^+\to f_0e^+\nu_{e}$& $\Koev$& $\Kstoev$   \\
			\hline
			CQM   ~\cite{Melikhov:2000yu}         & 2.48 & 0.92                 & 2.52         &               ...             &0.30         &    ...          \\
			RQM   ~\cite{RQM:2020}                & 2.37 & 0.87                 &2.69          & ...                           &0.40         &0.21           \\
			$\chi^{\rm UA}(I)$~\cite{chisq:2005}  & 1.7  & 0.74                 &   ...          &         ...                   &0.32         &  ...            \\
			$\chi^{\rm UA}(II)$~\cite{chisq:2005} & 2.5  & 0.61                 &  ...           & ...                          &0.2          &  ...             \\
			LCSR ~\cite{Verma:2011yw}             & 3.15$\pm$0.97&0.97$\pm$0.38 &...              &                ...           &    ...        &       ...        \\
			LFQM(I)~\cite{Wei:2009nc}             & 2.42 & 0.95                 &2.95          & ...                           &   ...         &        ...       \\
			LFQM(II)~\cite{Wei:2009nc}            &2.25  &0.91                  &2.58          &                     ...      &     ...       &         ...      \\
			LCSR~\cite{Offen:2013nma}             &2.00$\pm$0.32& 0.75$\pm$0.23 &         ...     &                ...            &    ...        &          ...     \\
			QM~\cite{Soni:2018adu}                &2.24& 0.83                   &3.01          &            ...               &0.20         &  ...             \\
			LCSR ~\cite{Hu:2021zmy}               &2.35$\pm$0.37 &0.79$\pm$0.13 &             ... &    ...                     &           ...  &      ...      \\
			LFQM ~\cite{H:2017}                   & ...&...                           & 2.9$\pm$0.3  &               ...            &0.27$\pm$0.02&0.19$\pm$0.02  \\
			LCSR~\cite{X:2021}                    &... &...                         & 2.46$\pm$0.42&          ...                  &0.39$\pm$0.08&0.23$\pm$0.03  \\
			LCSR ~\cite{Y:2006}                   & ...&                   ...        & 2.53$\pm$0.39&                            &0.39$\pm$0.07&0.23$\pm$0.03  \\
			CCQM ~\cite{Soni2020}                 &... &     ...                      &     ...         & 0.21$\pm$0.02              &     ...      &  ...             \\
			LCSR ~\cite{LCSR2}                    &... & ...                          & ...             &0.15$\pm$0.04               &    ...        & ...              \\
			LCSR ~\cite{LCSR1}                    & ...& ...                         & ...             &0.20$\pm$0.05               &       ...     & ...             \\
				\hline
			\hline
		\end{tabular}
	}
\end{table*}

\section{BESIII detector and Monte Carlo simulation}

The BESIII detector is a magnetic spectrometer~\cite{bes3} located at
the Beijing Electron Positron Collider
(BEPCII)~\cite{Yu:IPAC2016-TUYA01}. The cylindrical core of the BESIII
detector consists of a helium-based multilayer drift chamber (${\rm
  MDC}$), a plastic scintillator time-of-flight system (${\rm TOF}$),
  and a CsI(Tl) electromagnetic calorimeter (${\rm EMC}$), which are
  all enclosed in a superconducting solenoidal magnet providing a
  1.0~T magnetic field. The solenoid is supported by an octagonal
  flux-return yoke with resistive plate counter muon-identifier
  modules interleaved with steel. The acceptance of charged particles
  and photons is 93\% over the $4\pi$ solid angle. The
  charged-particle momentum resolution at $1~{\rm GeV}/c$ is $0.5\%$,
  and the resolution of specific ionization energy loss~(d$E$/d$x$) is
  $6\%$ for electrons from Bhabha scattering. The EMC measures photon
  energies with a resolution of $2.5\%$ ($5\%$) at $1$~GeV in the
  barrel (end-cap) region. The time resolution of the TOF barrel part
  is 68~ps, while that of the end-cap part was 110~ps. The end-cap TOF
  system was upgraded in 2015 using multi-gap resistive plate chamber
  technology, providing a time resolution of 60 ps~\cite{60ps1,60ps2}
  and benefiting 74\% of the data used in this analysis.  Details
  about the design and performance of the BESIII detector are given in
  Ref.~\cite{bes3}.

Simulated samples produced with {\sc geant4}-based~\cite{geant4} Monte
Carlo (MC) software, which includes the geometric description of the
BESIII detector and the detector response, are used to determine the
detection efficiency and to estimate backgrounds. The simulation
includes the beam-energy spread and initial-state radiation in
$e^+e^-$ annihilations modeled with the generator {\sc
  kkmc}~\cite{kkmc}.  Inclusive MC samples with luminosities of 20
times that of the data are produced at center-of-mass energies between
4.237 and 4.699 GeV. They include open-charm processes, initial state
radiation production of $\psi(3770)$, $\psi(3686)$ and $J/\psi$,
$q\bar q$ $(q=u, d, s)$ continuum processes, Bhabha scattering,
$e^+e^-\to\mu^+\mu^-$, $e^+e^-\to\tau^+\tau^-$, and
$e^+e^-\to\gamma\gamma$ events. In the simulation, the production of
open-charm processes directly via $e^+e^-$ annihilations is modeled
with the generator {\sc conexc}~\cite{conexc}.  The known decay modes
are modeled with {\sc evtgen}~\cite{evtgen} using branching
fractions taken from the PDG~\cite{pdg2024}, and the remaining unknown
decays of the charmonium states are modeled by {\sc
  lundcharm}~\cite{lundcharm}. Final-state radiation is incorporated
using {\sc photos}~\cite{photos}.  The input Born cross section line
shape of $e^+e^-\to D^{*+}_sD^{*-}_s$ is based on the results in
Ref.~\cite{crsDsDs}.  The input hadronic form factors for $\etaev$,
$\etapev$, $\phiev$, $\foev$, $\Koev$, and $\Kstoev$ are taken from
Refs.~\cite{Etaev,phimuv,Kev}.
\section{Analysis method}

In the $e^+e^-\to D^{*+}_{s} D^{*-}_{s}$ process, the $D_s^*$ mesons
will decay via $D_s^{*\pm}\to \gamma(\pi^0)D^\pm_s$.  As the first
step, we fully reconstruct a $D_s^{*-}$ meson in one of the chosen
hadronic decay modes, called a single-tag (ST) candidate, and then
attempt a reconstruction of a signal decay of the $D_s^{*+}$ meson. An
event containing both a ST and a signal decay is named a double-tag
(DT) candidate.  The branching fraction of the signal decay is
determined by \begin{equation}
	\mathcal B_{\rm sig}=\frac{N_{\rm DT}}{N_{\rm ST} \cdot \bar\epsilon_{{\rm sig}} \cdot{\mathcal B}_{\text{sub}}}.
	\label{eq1}
\end{equation}
Here, $N_{\rm DT}=\Sigma_{i,j} N_{\rm DT}^{i,j}$ and $N_{\rm
  ST}=\Sigma_{i,j} N_{\rm ST}^{i,j}$ are the total DT and ST yields in
  data summing over the tag mode $i$ and the energy point $j$; $\bar
  \epsilon_{\rm sig}$ is the averaged efficiency of the signal decay,
  and estimated by $\bar \epsilon_{\rm sig} = \sum_{j} \left[
  \sum_{i}\left( \frac{N_{\rm ST}^{i,j}}{N_{\rm ST}^{j}} \cdot
  \frac{\epsilon_{\rm DT}^{i,j}}{\epsilon_{\rm ST}^{i,j}} \right)
  \cdot \frac{N_{\rm ST}^{j}}{N_{\rm ST}}\right]$, where
  $\epsilon_{\rm DT}^{i,j}$ and $\epsilon_{\rm ST}^{i,j}$ are the
  detection efficiencies of the DT and ST candidates for the $i$-th
  tag mode at the $j$-th energy point, respectively. $N_{\rm
  ST}^{i,j}$ and $N_{\rm ST}^{j}$ are the ST yields for the $i$-th tag
  mode at the $j$-th energy point and the total ST yield at the $j$-th
  energy point, respectively.  The efficiencies are estimated from MC
  samples and do not include the branching fractions of the sub-decay
  channels used for the signal and ST reconstruction. ${\mathcal B}_{\text{sub}}$ is the
  product of the branching fractions of the  intermediate decays  in the signal decay. 

\section{\boldmath Single-tag $D^{*-}_s$ candidates}

The ST $D_s^{*-}$ candidates are reconstructed via $D_s^{*-}\to
\gamma(\pi^0)D^-_s$, and the $D^-_s$ candidates are reconstructed in
the hadronic decay modes of $D^-_s\to K^+K^-\pi^-$,
$K^+K^-\pi^-\pi^0$, $K^0_SK^-$, $K^0_SK^-\pi^0$, $K^0_SK^0_S\pi^-$,
$K^0_SK^+\pi^-\pi^-$, $K^0_SK^-\pi^+\pi^-$, $\pi^+\pi^-\pi^-$,
$K^+\pi^-\pi^-$, $\eta_{\gamma\gamma}\pi^-$,
$\eta_{\pi^0\pi^+\pi^-}\pi^-$,
$\eta^\prime_{\eta_{\gamma\gamma}\pi^+\pi^-}\pi^-$,
$\eta^\prime_{\gamma\rho^0}\pi^-$, and $\eta_{\gamma\gamma}\rho^-$.
Throughout this paper, the subscripts of $\eta$ and $\eta^{\prime}$
denote the decay modes used to reconstruct $\eta$ and $\eta^{\prime}$,
respectively.

All charged tracks are required to be within $|\!\cos\theta|<0.93$,
where $\theta$ is the polar angle with respect to the $z$- axis, which
is the MDC symmetry axis.  Those not originating from $K^0_S$ decays
are required to satisfy $|V_{xy}|<1$ cm and $|V_{z}|<10$ cm, where
$|V_{xy}|$ and $|V_{z}|$ are distances of the closest approach to the
interaction point (IP) in the transverse plane and along the $z$-axis,
respectively.  The charged tracks are identified with a particle
identification (PID) procedure, in which both the $dE/dx$ and TOF
measurements are combined to form confidence levels for pion and kaon
hypotheses, e.g., $CL_\pi$ and $CL_K$.  Kaon and pion candidates are
required to satisfy $CL_{K}>CL_{\pi}$ and $CL_{\pi}>CL_{K}$,
respectively.

Candidates for $K_S^0$ are reconstructed via the decays  $K^0_S\to
\pi^+\pi^-$.  The distances of closest approach of the $\pi^\pm$
candidates to the IP must satisfy $|V_{z}|<20$ cm
without any $|V_{xy}|$ requirement.  No PID requirements are applied
for the two charged pions.  For any $K^0_S$ candidate, the
$\pi^+\pi^-$ invariant mass is required to be within $\pm$12 MeV/$c^2$
around the known $K^0_S$ mass~\cite{pdg2024}.  A secondary vertex fit
is performed, and the decay length must be greater than twice the
vertex resolution away from the IP.

Photon candidates are selected from shower clusters in the EMC.
The difference between the shower time and the event start time must
be within $[0,700]$~ns to remove showers unrelated to the event. This selection retains more than 99\% of reconstructed signal photons and removes 75\% of background energy depositions in the EMC. The
energy of each shower is required to be greater than 25~MeV in the
barrel EMC region and 50 MeV in the end-cap EMC region~\cite{bes3}.
To exclude showers originating from charged tracks, the opening angle
subtended by the EMC shower and the position of any charged track at
the EMC is required to be greater than 10 degrees as measured from the
IP.

Candidates for $\pi^0(\eta)$ are reconstructed via
$\pi^0(\eta)\to\gamma\gamma$ decays.  The $\gamma\gamma$ invariant
masses are required to be within $(0.115,0.150)$~GeV$/c^{2}$ and
$(0.500,0.570)$~GeV$/c^{2}$, respectively.  To improve the momentum
resolution and suppress background, a kinematic fit constraining the
$\gamma\gamma$ invariant mass to the $\pi^0(\eta)$ known
mass~\cite{pdg2024} is performed on the selected $\gamma\gamma$
pairs. The updated four-momenta of the photon pairs are used for
further analysis.

The $\eta$ candidates are also reconstructed via $\eta\to
\pi^+\pi^-\pi^0$ decays, in which the $\pi^+\pi^-\pi^0$ invariant
masses are required to lie in the mass window
$(0.53,0.57)~\mathrm{GeV}/c^2$.

The $\eta^\prime$ candidates are reconstructed via $\eta^\prime\to
\pi^+\pi^-\eta$ and $\eta^\prime \to \gamma\rho^0$ decays, and the
$\pi^+\pi^-\eta$ and $\gamma\rho^0$ invariant masses are required to
lie in the mass windows $(0.946,0.970)$ GeV/$c^2$ and
$(0.940,0.976)~\mathrm{GeV}/c^2$, respectively. For
$\eta^\prime\to\gamma\rho^0$, the minimum energy of the radiative
photon produced in the $\eta^\prime$ decays is required to be greater
than 0.1~GeV.

The $\rho^0$ and $\rho^+$ candidates are reconstructed from
$\rho^0\to \pi^+\pi^-$ and $\rho^+\to \pi^+\pi^0$ decays, in which the
$\pi^+\pi^{-(0)}$ invariant masses are required to be within
$(0.57,0.97)~\mathrm{GeV}/c^2$.

To suppress the contributions of $D^-_s\to K^0_S(\to \pi^+\pi^-)\pi^-$
and $D^-_s\to K^0_S(\to \pi^+\pi^-)K^-$ for the $D^-_s\to
\pi^+\pi^-\pi^-$ and $D^-_s\to K^+\pi^-\pi^-$ tag modes, we reject any
candidates with the $\pi^+\pi^-$ invariant mass being in the mass
window $(0.468,0.518)$~GeV/$c^2$.

The invariant masses of tagged $D^-_s$ candidates are required to be
within the mass windows according to Refs.~\cite{Etaev}.  To
further distinguish the single-tag $D^{*-}_s$ from combinatorial
background, we use two kinematic variables: the energy difference
defined as
\begin{equation}
	\Delta E =  E_{\rm beam}-E_{\rm tag},
	\label{eq:deltaE}
\end{equation}
and the beam constrained mass
\begin{equation}
M_{\rm BC}=\sqrt{E^2_{\rm beam}/c^4-|\vec{p}_{\rm tag}|^2/c^2}.
\label{eq:mBC}
\end{equation}
 Here $E_{\rm beam}$ denotes the beam energy, while $E_{\rm tag}$
 and $\vec p_{\rm tag}$ are respectively the energy and
  momentum of the ST $D_s^{*-}$ candidate in the rest frame
 of the initial $e^+e^-$ beams.  The correctly reconstructed ST
 candidates are expected to peak around zero and the known $D_s^{*}$
 mass in the $\Delta E$ and $M_{\rm BC}$ distributions, respectively.
At a given energy point, we choose the same $M_{\rm BC}$  signal regions for different tag modes due to similar resolutions, while the $M_{\rm BC}$ signal regions slightly expand with energy.

\begin{figure*}[htbp]
\centering
	\includegraphics[width=0.8\textwidth] {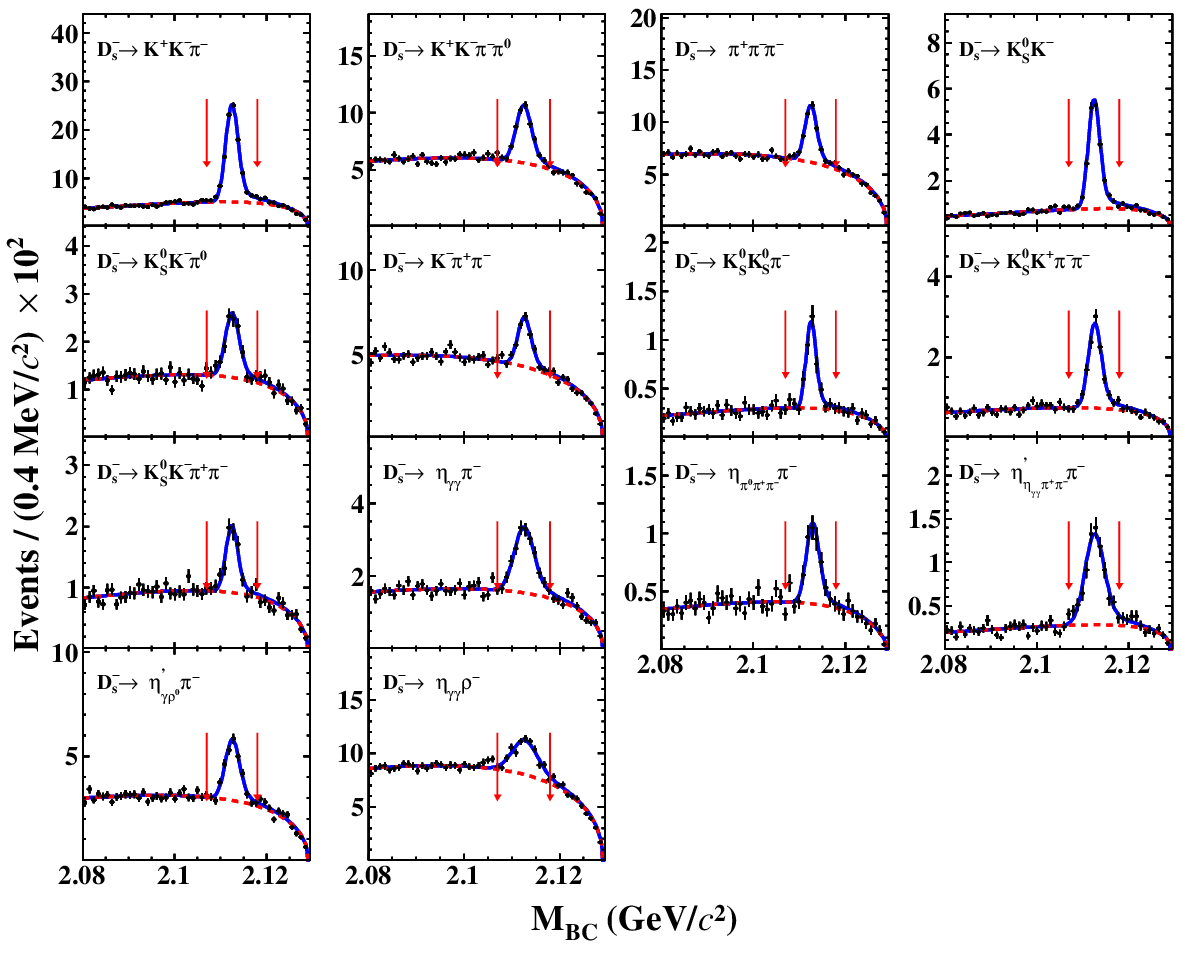}
\caption{\footnotesize
Fits to the $M_{\rm BC}$ distributions of the ST $D^{*-}_s$ candidates, where the points with error bars are data at 4.260 GeV, the solid curves show the best fits, and the red  dashed curves show the fitted   combinatiorial  background  shapes. The pairs of arrows denote the $M_{\rm BC}$ signal window.}
\label{fig:stfit}
\end{figure*}

For each tag mode, the candidate giving the minimum $|\Delta E|$ value
is chosen if there are multiple $\gamma/\pi^0$ or $D_s$ combinations
in an event. Table~\ref{tab:ST} shows the mass windows for $D^-_s$ and  the
$\Delta E$ requirements for $D^{*-}_s$.   The resultant $M_{\rm BC}$ distributions of the accepted
ST candidates of different tag modes at 4.260 GeV are shown in
Fig.~\ref{fig:stfit}. Similar distributions are also obtained at the
other center-of-mass energy points.  The yields of the ST $D_s^{*-}$
mesons are obtained from fits to the individual $M_{\rm BC}$ spectra.
The fits are performed to each of the data sets taken below 4.5 GeV
due to the relatively large samples.  The data samples taken above 4.5
GeV are combined into one data set due to the limited number of
events.  For all fits, the signals are described by the MC-simulated
shape convolved with a Gaussian function to account for the resolution
difference between data and MC simulation. The range of the mean value of the convolved normal distribution is $(-0.002, 0.002)$~GeV/$c^2$, with a resolution range of $(0,0.005)$~GeV/$c^2$. For each data set taken
below 4.5 GeV, the combinatorial background is described by an ARGUS
function~\cite{argus}, while for the combined data set above 4.5 GeV,
the combinatorial background is described by a cubic  polynomial
function, which has been validated with the inclusive MC sample.
Figure~\ref{fig:stfit} also shows the results of the fits to the
$M_{\rm BC}$ distributions of the ST $D_s^{*-}$ candidates at 4.260
GeV. The candidates in the $M_{\rm BC}$ signal regions, indicated by
the red arrows in each sub-figure, are retained for the further
analysis.   The obtained ST yields in data ($N_{\rm
	ST}^{i}$) and the    ST efficiencies ($\epsilon_{\rm ST}^{i}$) for
different tag modes are also shown in Table~\ref{tab:ST}.  Table ~\ref{tab:sigeff} shows the $M_{\rm BC}$
signal regions and the total ST yields at the different energy points.
The total ST yield in data is $N^{\rm tot}_{\rm ST}$ = 124027$\pm$1121.

\begin{table*}[htbp]
\caption{The mass windows for $D^-_s$~\cite{Etaev}, the $\Delta E$ requirements for $D^{*-}_s$, the ST yields in data and the ST efficiencies at 4.260 GeV, where the efficiencies do not
	include the branching fractions for the sub-resonant decays and the uncertainties are statistical only.   \label{tab:ST}}
\centering
\begin{tabular}{lccr@{}l r@{}l}
\hline\hline
$D^-_s$ tag mode &$M_{D^-_s}$ (GeV/$c^2$)&$\Delta E$ (MeV) &\multicolumn{2}{c}{$N_{\rm ST}$} &\multicolumn{2}{c}{\makecell{$\epsilon_{\rm ST}$(\%)}}\\
\hline
$K^{+} K^{-}\pi^{-}$                          &$(1.950,1.986)$&$(-26, 31)$&$7454$ &$\pm125$&$19.67$&$\pm0.07$\\
$K^{+} K^{-}\pi^{-}\pi^{0}$                   &$(1.947,1.982)$&$(-29, 38)$&$2186$ &$\pm108$& $5.18$&$\pm0.06$ \\
$\pi^{+}\pi^{-}\pi^{-}$                       &$(1.952,1.984)$&$(-28, 34)$&$1929$ &$\pm99$&$26.20$&$\pm0.26$\\
$K_S^{0} K^{-}$                               &$(1.948,1.991)$&$(-30, 33)$&$1649$ &$\pm53$&$22.83$&$\pm0.16$ \\
$K_S^{0} K^{-}\pi^{0}$                        &$(1.946,1.987)$&$(-31, 40)$&$554 $ &$\pm50$&$6.99 $&$\pm0.12$\\
$K^{-}\pi^{+}\pi^{-}$                         &$(1.953,1.983)$&$(-28, 33)$&$1112$ &$\pm83$&$22.84$&$\pm0.38$ \\
$K_S^{0} K_S^{0}\pi^{-}$                      &$(1.951,1.986)$&$(-28, 32)$&$266 $ &$\pm22 $&$11.50$&$\pm0.21$\\
$K_S^{0} K^{+}\pi^{-}\pi^{-}$                 &$(1.953,1.983)$&$(-26, 31)$&$808 $ &$\pm45 $&$9.68 $&$\pm0.11$ \\
$K_S^{0} K^{-}\pi^{+}\pi^{-}$                 &$(1.958,1.980)$&$(-26, 31)$&$390 $ &$\pm40 $&$9.19 $&$\pm0.19$\\
$\eta_{\gamma\gamma}\pi^{-}$                  &$(1.930,2.000)$&$(-43, 52)$&$983 $ &$\pm69 $&$19.19$&$\pm0.28$ \\
$\eta_{\pi^0\pi^{+}\pi^{-}}\pi^{-}$           &$(1.941,1.990)$&$(-34, 43)$&$269 $ &$\pm29 $&$11.71$&$\pm0.28$\\
$\eta^{\prime}_{\eta\pi^{+}\pi^{-}} \pi^{-}$     &$(1.940,1.996)$&$(-34, 40)$&$575 $ &$\pm40 $&$11.49$&$\pm0.18$ \\
$\eta^{\prime}_{\gamma\rho^{0}} \pi^{-}$         &$(1.938,1.992)$&$(-33, 43)$&$1233$ &$\pm75$&$14.11$&$\pm0.19$\\
$\eta_{\gamma\gamma}\rho^{-}_{\pi^{-}\pi^{0}}$&$(1.920,2.006)$&$(-49, 66)$&$2142$ &$\pm191$&$7.93 $&$\pm0.13$  \\
\hline\hline
\end{tabular}
\end{table*}

\begin{table}[htbp]\centering
	\caption{The integrated luminosities (${\mathcal L}$), $M_{\rm BC}$ requirements, and  ST yields in data ($N_{\rm ST}$) for various energy points. The uncertainties are statistical only.}
	\label{tab:sigeff}
		\begin{tabular}{lcc r@{}l}
			\hline
			\hline
			$E_{\rm cm}$ (GeV)&${\mathcal L}$ (pb$^{-1}$) &$M_{\rm BC}$ (GeV/$c^2$) &\multicolumn{2}{c}{\makecell{ $N_{\rm ST}$ }}\\
			\hline
			4.237       &530.3&$(2.107,2.117)$&$6477$&$\pm163$    \\
			4.246       &593.9&$(2.107,2.118)$&$11944$&$\pm246$    \\
			4.260       &828.4&$(2.107,2.118)$&$21550$&$\pm320$    \\
			4.270       &531.1&$(2.107,2.118)$&$13319$&$\pm244$   \\
			4.280       &175.7&$(2.106,2.119)$&$4063$&$\pm152$    \\
			4.290       &502.4&$(2.106,2.119)$&$9316$&$\pm221$    \\
			4.310-4.315       &546.3&$(2.106,2.119)$&$5758$&$\pm228$    \\
			4.400       &507.8&$(2.106,2.119)$&$1855$&$\pm87$     \\
			4.420       &1090.7&$(2.106,2.121)$&$14890$&$\pm443$   \\
			4.440       &569.9&$(2.106,2.121)$&$9699$&$\pm443$    \\
			4.470-4.699 &4768.3&$(2.104,2.123)$&$25156$&$\pm762$   \\

			\hline\hline
		\end{tabular}
	\end{table}

\section{Double-tag events}
At the recoil sides of the ST $D_s^{*-}$ mesons, the radiative photons
or  $\pi^0$ of the $D_s^{*+}$ decays and the candidates for semileptonic $D^+_s$ decays are
selected with the surviving neutral and charged tracks which have not
been used in the ST selection.

The candidates for $\gamma$, $\pi^0$, $\pi^\pm$, $K^\pm$, $K^0_S$,
$\rho^+$, $\eta$, and $\eta^\prime$ are selected with the same
selection criteria as those used on the tag side.  The $K^{*0}$,
$f_0$, and $\phi$ candidates are reconstructed with the decays
$K^{*0}\to K^+\pi^-$, $f_0\to \pi^+\pi^-$, and $\phi\to K^+K^-$,
respectively, and their invariant masses are required to be within
$(0.882, 0.992)$ GeV$/c^2$, $(0.880, 1.080)$ GeV$/c^2$, and $(1.004,
1.034)$ GeV$/c^2$, respectively.

The $e^+$ candidates are identified by using the $dE/dx$, TOF, and EMC
information.  Confidence levels for the pion, kaon and positron
hypotheses ($CL_\pi$, $CL_K$ and $CL_e$) are formed.  Charged tracks
satisfying $CL_e>0.001$ and $CL_e/(CL_e+CL_\pi+CL_K)>0.8$ are assigned
as $e^+$ candidates.  The energy loss of the positron due to
bremsstrahlung is partially recovered by adding the energies of the
EMC showers that are within $10^\circ$ of the positron direction at
the IP and not matched to other particles.

Signal decay candidates are required to
have no extra charged tracks to suppress hadronic related background
events.  To suppress the backgrounds with an extra photon(s), the maximum
energy of showers which have not been used in the DT selection,
denoted as $E_{\mathrm{extra}~\gamma}^{\rm max}$, is required to be
less than 0.3~GeV.

For the semileptonic $D^+_s$ decays, the invariant masses of the
hadron and lepton of the signal side are required to be less than 1.90
GeV/$c^2$ for the Cabibbo favored decays and to be less than 1.75
GeV/$c^2$ for the Cabibbo suppressed decays in order to minimize
hadronic $D_s$ backgrounds.

To separate signal from combinatorial background, we define the
missing mass squared of the undetectable neutrino(s) by
\begin{equation}
M^2_{\mathrm{miss}}\equiv E^2_{\mathrm{miss}}/c^4-|\vec{p}_{\mathrm{miss}}|^2/c^2.
\end{equation}
Here, $E_{\mathrm{miss}}\equiv
E_{\mathrm{beam}}-E_{\gamma(\pi^0)}-E_{h}-E_{\ell}$ and
$\vec{p}_{\mathrm{miss}}\equiv
\vec{p}_{D^{*+}_s}-\vec{p}_{\gamma(\pi^0)}-\vec{p}_{h}-\vec{p}_{\ell}$
are the missing energy and momentum of the DT event in the $e^+e^-$
center-of-mass frame, in which $E_{i}$ and $\vec{p}_{i}$
($i=\gamma(\pi^0)$, $h$ or $\ell$) are the energy and momentum of the
$i$ particle in the recoil side.  The $M^2_{\rm miss}$ resolution is
improved by constraining the $D^{*+}_s$ energy to the beam energy and
\begin{equation}
 \vec{p}_{D^{*+}_s} \equiv
{-\hat{p}_{D^{*-}_s}}\cdot\sqrt{E_{\mathrm{beam}}^{2}/c^2-m_{D^{*}_s}^{2}c^2}, 
\end{equation}
where $\hat{p}_{D^{*-}_s}$ is the unit vector in the momentum
direction of the ST $D^{*-}_s$ and $m_{D^{*}_s}$ is the known
$D^{*}_s$ mass~\cite{pdg2024}.  For the correctly reconstructed signal
events of the semileptonic $D^{+}_s$ decays, the $M^2_{\rm miss}$
distributions are expected to peak around zero.

\begin{figure*}[htbp]
	\centering
	\includegraphics[width=0.8\linewidth]{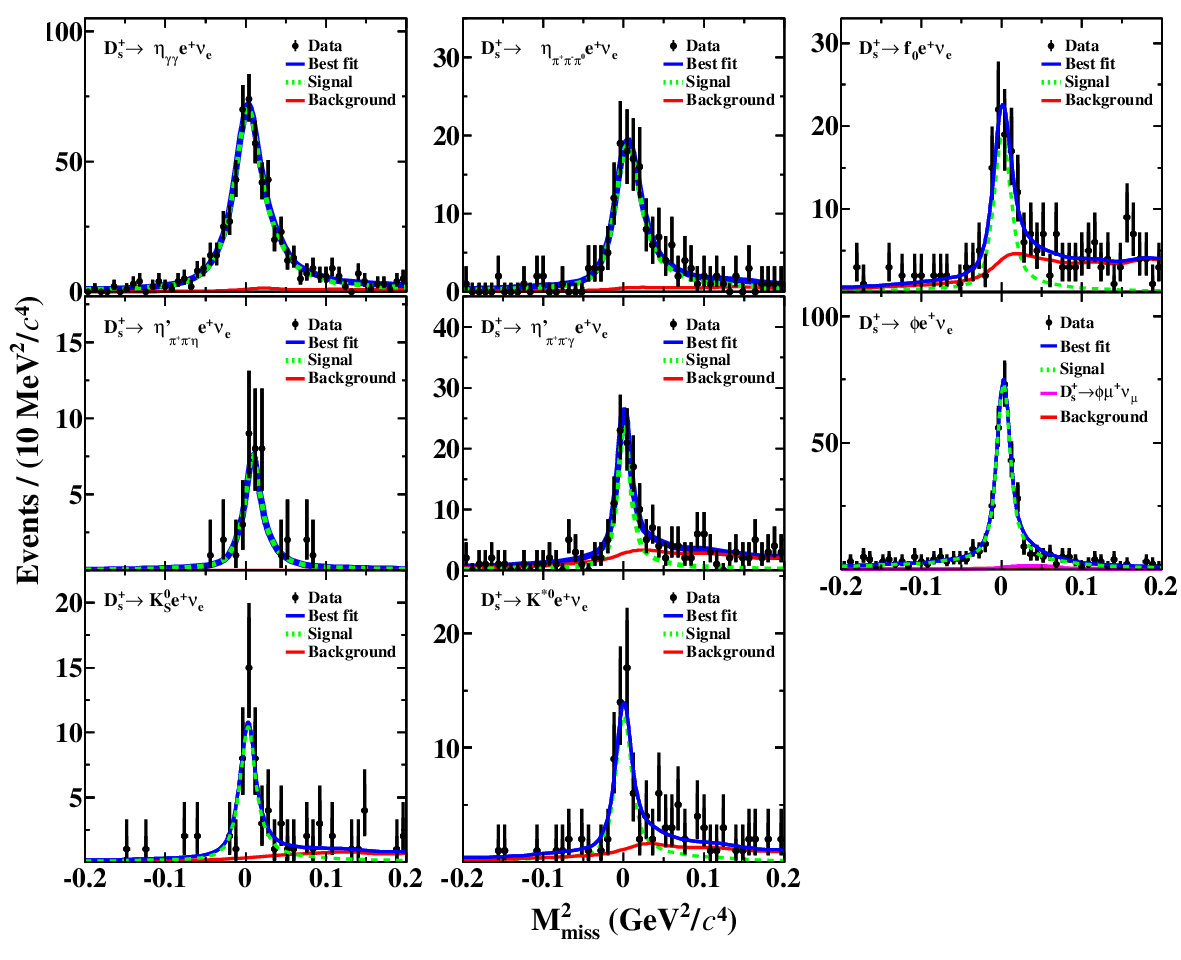}
	\caption{Fits to the $M^{2}_{\rm miss}$ distributions of the candidate events for the semileptonic $D^+_s$ decays. The points with error bars represent the  data. The blue solid curves denote the best fits. The green dotted curves and red solid curves show the fitted signal shape and combinatorial background shape. For $D_s^+ \to \phi e^+ \nu_{e}$, the purple solid curve is the peaking background from $D_s^+ \to \phi \mu^+ \nu_{\mu}$. \label{fig:semi}}
\end{figure*}

Figure~\ref{fig:semi} shows the resulting $M^{2}_{\rm miss}$
distributions of the accepted candidate events for the semileptonic
$D^+_s$ decays for DT events from all energy points.  The yields of different signal decays are obtained
from unbinned maximum likelihood fits to these distributions.  In the fits, the signal  is modeled by the simulated shape extracted from the signal MC sample, and 
the background is modeled by the simulated shape derived from the  inclusive MC sample.
To compensate the difference in resolutions between data and MC simulation, the simulated signal shape is convolved with a normal function with free parameters. The size and shapes of the peaking background of $D^+_s \to \phi \mu^+ \nu_{\mu}$ for $D^+_s \to \phi e^+ \nu_{e}$ are fixed based on MC simulation; while  the muon related background for other decays is  included in the combinatorial background due to relatively less contribution.  
For $D^+_s\to \eta
e^+\nu_e$ or $D^+_s\to \eta^\prime e^+\nu_e$ decays, simultaneous fits
are performed on the distributions of the accepted candidates
reconstructed in the two $\eta/\eta^{\prime}$ decay modes, in which
they are constrained to share a common branching fraction after taking
into account the differences of signal efficiencies and branching
fractions between the two decay modes. For these two decays, their   signal yields are estimated by Eq.~\ref{eq1}, and both  $\mathcal B_{\rm sig}$ and background yields are left free. For $\phiev$, $\foev$,
$\Koev$ and $\Kstoev$, the yields of signal and combinatorial backgrounds are free. 
Table \ref{tab:lep} summarizes
the detection efficiencies, the signal yields, and the measured
branching fractions of different semileptonic $D^+_s$ decays.  It
should be noted that the listed branching fraction of $D^+_s\to f_0
e^+\nu_e$ has not been normalized by the branching fraction of $f_0\to
\pi^+\pi^-$ because it is not well known.  Previous studies via $e^+e^- \to D_sD_s^*$ with
higher statistics show that the non-resonant components in the decays
 $D^+_s \to \eta^{(\prime)}\ell^+\nu_\ell$~\cite{Etaev,panx1},
$D^+_s \to \phi \mu^+ \nu_{\mu}$~\cite{phimuv}, $D^+_s \to
f_0e^+\nu_{e}$\cite{f0ev} and $D^+_s \to K^{(*)0}e^+\nu_e$~\cite{Kev}
are negligible, therefore they are ignored in this work.

\begin{table*}
\centering
\caption{Signal efficiencies ($\epsilon_{\rm sig}$), signal yields ($N_{\rm DT}$), products of branching fractions of the intermediate decays in the signal decay ($\mathcal B_{\rm sub}$),  and measured branching
  fractions ($\mathcal B_{\rm sig}$) for various signal decays. For $\epsilon_{\rm
    sig}$ and $N_{\rm DT}$, the uncertainties are statistical only;
  for $\mathcal B_{\rm sig}$, the first and second
  uncertainties are statistical and systematic, respectively.  It
  should be noted that the listed branching fraction of $D^+_s\to
  f_0 e^+\nu_e$ has not been normalized by the branching fraction of
  $f_0\to \pi^+\pi^-$ because it is not well known.
\label{tab:lep}}
	\begin{tabular}{lcccc}
	\hline\hline
Signal decay       & $\epsilon_{\rm sig}$ (\%)   & $\mathcal B_{\rm sub}$ (\%)& $N_{\rm DT}$                      & $\mathcal B_{\rm sig}$ (\%)  \\ \hline
$D_{s}^{+} \to \eta_{\gamma \gamma} e^{+} \nu_{e}$                   & $50.78\pm0.12$ & $39.36\pm0.18$       & \multirow{2}{*}{$716.2\pm33.8$} & \multirow{2}{*}{$2.35\pm 0.11\pm0.10$}\\
$D_{s}^{+} \to \eta_{\pi^{+} \pi^{-} \pi^{0}} e^{+} \nu_{e}$         & $20.42\pm0.08$ &    $32.18\pm0.07$    &                                &                                    \\
$D_{s}^{+} \to \eta^{\prime}_{\pi^{+} \pi^{-} \eta} e^{+} \nu_{e}$   & $22.35\pm0.07$ &  $16.72\pm0.30$      & \multirow{2}{*}{$133.7 \pm 14.5 $} & \multirow{2}{*}{$0.82\pm0.09\pm0.04$} \\
$D_{s}^{+} \to \eta^{\prime}_{\pi^{+} \pi^{-} \gamma} e^{+} \nu_{e}$ & $32.48\pm0.09$ &     $29.50\pm0.40$   &                                &                                    \\
$D_{s}^{+} \to \phi_{K^{+} K^{-}} e^{+} \nu_{e}$                     & $25.48\pm0.07$ &    $49.10\pm0.50$    & $350.2\pm24.5$                  & $2.21\pm0.16\pm0.11$                  \\
$D_{s}^{+} \to f_{0_{\pi^+\pi^-}} e^{+} \nu_{e}$        &$46.24\pm0.11$ &   ...    & $91.0\pm14.1$                   & $0.15\pm0.02\pm0.01$                  \\
$D_{s}^{+} \to K^{0} e^{+} \nu_{e}$                                  & $46.21\pm0.11$  &  $34.60\pm0.03$    & $50.5\pm 8.4$                   & $0.24\pm0.04\pm0.01$                  \\
$D_{s}^{+} \to K^{0\star} e^{+} \nu_{e}$                             & $41.78\pm0.10$  & $66.67$      & $65.4\pm 10.9$                   & $0.19\pm0.03\pm0.01$                  \\

	\hline\hline
\end{tabular}
\end{table*}

\section{Systematic uncertainties}

With the DT method, most systematic uncertainties related to the ST
selection cancel. Details about the systematic uncertainties in the
measurements of the branching fractions of semileptonic $D_S^+$
decays are discussed below.  Table~\ref{sys} summarizes the sources of
the systematic uncertainties in the measurements of the branching
fractions of $D_s^+\to \eta^{(\prime)}e^+\nu_e$, $\phiev$, $\foev$,
$\Koev$ and $\Kstoev$. They are assigned relative to the measured
branching fractions. 
For $D_s^+\to \eta^{(\prime)} e^+\nu_e$, the systematic
uncertainties due to $N^{\rm tot}_{\rm ST}$, $\gamma/\pi^0/\eta\to\gamma\gamma$ reconstruction, $e^\pm(\pi^\pm)$ tracking/PID, kinematic fit, $E_{\rm extra \gamma}^{\rm max}$ and $N_{\rm char}^{\rm extra}$,  as well as the simultaneous fit to $M^2_{\rm miss}$  are correlated, and two $\eta/\eta^\prime$ decay modes share a common value for each correlated source in Table~\ref{sys}.  The remaining uncertainties are uncorrelated, and the  two $\eta/\eta^\prime$ decay modes have individual values for each uncorrelated  source in Table~\ref{sys}. 

The total systematic uncertainties of the branching fractions of
$D_s^+\to \eta e^+\nu_e$ and $D_s^+\to \eta^\prime e^+\nu_e$ are 4.5\%
and 5.3\%, respectively, after taking into account correlated and
uncorrelated systematic uncertainties and using the method described
in Ref.~\cite{Schmelling:1994pz}.  The total systematic uncertainties
in the measurements of the branching fractions of $\phiev$, $\foev$,
$\Koev$ and $\Kstoev$ are 4.8\%, 5.4\%, 6.1\%, and 5.2\%, by adding
the individual uncertainties in quadrature.

\subsection{\boldmath Number of ST $D^{*-}_s$ events}

The systematic uncertainty in the $M_{\rm BC}$ fits is estimated by
using alternative signal and background shapes, and repeating the fit
for both data and the inclusive MC sample.  For an alternative signal
shape, we require, in addition to all other requirements, that the
reconstructed $\gamma(\pi^0)$ and $D^{*-}_s$ agree within $20^{\circ}$
of the generated ones.  
For each data set below 4.5 GeV, the background shape is changed to a third-order Chebyshev polynomial, while for data set above 4.5 GeV, the background shape is changed to a fourth-order Chebyshev polynomial.
The relative difference of the ST yields
is assigned as the systematic uncertainty.  In addition, the
uncertainty due to the fluctuation of the fitted ST yield is
considered as another systematic uncertainty, since it affects the
selection of the DT events.  The quadrature sum of these two items,
1.9\%, is assigned as the corresponding systematic uncertainty.

\subsection{Tracking and PID}

The tracking and PID efficiencies of $\pi^\pm$ and $K^\pm$ were
studied with control samples of $e^+e^-\to K^+K^-\pi^+\pi^-$.  The
efficiencies of tracking of $e^+$ were studied with a control sample
of Bhabha scattering events of $e^+e^-\to \gamma e^{+} e^{-}$.  The
systematic uncertainty for both tracking and PID efficiency of
$\pi^\pm$, $K^\pm$, and $e^+$ is assigned to be 1.0\% per charged
track.

\subsection{\boldmath $K^0_S$ reconstruction}
The systematic uncertainty in the $K_{S}^{0}$ reconstruction
efficiency is estimated with $J/\psi\to
K^{*\mp}K^{\pm}$ and $J/\psi\to \phi K_S^{0}K^{\pm}\pi^{\mp}$ control
samples~\cite{sysks} and found to be 1.5\% per $K^0_S$.

\subsection{\boldmath Selection of $\gamma$, $\pi^0$, and $\eta$}

The systematic uncertainty in the transition $\gamma$ reconstruction
is 1.0\% according to Ref.~\cite{sysgamma}.  The systematic
uncertainty in the $\pi^0$ reconstruction was studied by using a
sample of $e^+e^-\to K^+K^-\pi^+\pi^-\pi^0$, and  the systematic
uncertainty is 1.0\% for each $\pi^0$.
The systematic uncertainty in the $\eta \to \gamma\gamma$ reconstruction
is assumed to be 1.0\%, the same as $\pi^{0}$ due to limited $\eta$
events.  If there are $\gamma$, $\pi^0$, and $\eta$ combinations, the
total systematic uncertainty is added linearly to be conservative.

\subsection{\boldmath Mass windows of $\eta_{\pi^+\pi^-\pi^0}$, $\eta^\prime$,  $\phi$, $f_0$, and $K^{*0}$}
The systematic uncertainties due to the mass windows of
$M_{\pi^0\pi^+\pi^-}$, $M_{\eta\pi^+\pi^-}$, and
$M_{\pi^+\pi^-\gamma}$ are assigned as 0.1\%, 0.1\%, and 1.0\%,
respectively, using the control samples of
$J/\psi\to\phi\eta^{(\prime)}$~\cite{Etaev}.  The systematic
uncertainties in the requirements of $M_{K^+\pi^-}$, $M_{\pi^+\pi^-}$,
and $M_{K^+K^-}$, are studied with the control samples of $D^+\to
K^{*0}e^+\nu_{e}$, $D^+_{s}\to f_0e^+\nu_{e}$, and $D^0 \to
K^0_{S}\phi$, and the differences of the efficiencies of each
mass window between data and MC simulation, 1.2\%, 0.2\% and 0.2\%,
respectively, are taken as their systematic uncertainties.  The
efficiencies of the requirements of the invariant masses of the hadron
and lepton of the signal side are greater than 99\% for all signal
decays, and the differences of these efficiencies between data and MC
simulation are negligible.

\subsection{Kinematic fit}

The systematic uncertainty due to the kinematic fit is studied by
using control samples of $D_s^+\to K^+K^-\pi^+$ and $D_s^+\to
\eta\pi^0\pi^+$. The larger difference of the acceptance efficiencies
between data and MC simulation is taken as the corresponding
systematic uncertainty.

\subsection{MC statistics and MC model}

The uncertainty due to the limited MC statistics is considered as a
source of systematic uncertainty.  The systematic uncertainties due to
the MC model are examined by varying the input hadronic form factors
by $\pm 1\sigma$.  The changes of the signal efficiencies are taken as
the systematic uncertainties.

\subsection{Quoted branching fractions}

The uncertainties in the quoted branching fractions are from
$\eta\to\gamma\gamma$, $\eta\to\tpi$ $\eta^\prime\to\pi^+\pi^-\eta$,
$\eta^\prime\to \pi^+\pi^-\gamma$, $D_s^{*-}\to\gamma(\pi^0) D_s^+$,
$\pi^0\to\gamma\gamma$, $K^0_S\to \pi^+\pi^-$, and $\phi\to
K^+K^-$~\cite{pdg2024}.  The quoted branching fractions are $\mathcal
B(\pi^0\to\gamma\gamma)=(98.823\pm0.034)\%$, $\mathcal
B(\eta\to\gamma\gamma)=(39.41\pm0.20)\%$, $\mathcal
B(\eta\to\tpi)=(22.92\pm0.28)\%$, $\mathcal
B(\eta^\prime\to\pi^+\pi^-\eta)=(42.5\pm0.5)\%$, $\mathcal
B(\eta^\prime\to\pi^+\pi^-\gamma)=(29.5\pm0.4)\%$, $\mathcal
B(K^0_S\to \pi^+\pi^-)=(69.20\pm0.05)\%$, and $\mathcal B(\phi\to
K^+K^-)=(49.1\pm0.5)\%$. Their uncertainties, 0.1\%, 0.5\%, 1.2\%,
1.2\%, 1.4\%, 0.07\%, and 1.1\%, are taken as the systematic
uncertainties.

\subsection{\boldmath $M^2_{\rm miss}$ fit}

The systematic uncertainty of the $M^2_{\rm miss}$ fit is
determined by varying the signal and background shapes.  The
uncertainty in the signal shape is estimated by replacing the nominal
shape with the simulated shape convolved with  a sum of two normal distributions with floating parameters.  The systematic uncertainty caused
by the background shape is considered in three ways. First, we use
alternative MC-simulated shapes by varying the relative fractions of
the main backgrounds from $D_s^\pm D_s^{*\mp}$, $D_s^{*+}D_s^{*-}$,
open charm and $q\bar q$ by $\pm 1\sigma$ of individual observed cross
sections~\cite{crsDsDs}. Second, we use a straight line  for the background.  Third, we vary the yields of
the main background sources by $\pm 1\sigma$ of the quoted branching
fractions~\cite{pdg2024}.  The changes of the re-measured branching
fractions are assigned as the corresponding systematic uncertainties.
For each signal decay, the total systematic uncertainty is assigned as
the quadratic sum of the effects mentioned in this subsection.

\begin{table*}[htp]
	\centering
	\caption{Relative systematic uncertainties (in \%) in the branching
		fraction measurements.  
		The top parts of systematic uncertainties are correlated and the bottom  parts are
		uncorrelated for $\etaev$ and
		$\etapev$.
		\label{sys}}
	\begin{tabular}{l|cc|cc|c|c|c|c}
		\hline
		\hline
		Source  & $\eta_{\gamma\gamma}e^+\nu_e$&$\eta_{\pi^+\pi^-\pi^0}e^+\nu_e$&$\eta^\prime_{\pi^+\pi^-\eta}e^+\nu_e$&$\eta^\prime_{\gamma\rho^0}e^+\nu_e$&$\phi e^+\nu_e$&$f_0e^+\nu_e$&$K^0e^+\nu_e$&$K^*e^+\nu_e$  \\
		\hline
		$N_{\rm ST}$            &\multicolumn{2}{c|}{1.9} &\multicolumn{2}{c|}{1.9}&1.9 &1.9 &1.9 &1.9 \\
		$\gamma/\pi^0/\eta\to\gamma\gamma$ reconstruction&\multicolumn{2}{c|}{2.0}  &\multicolumn{2}{c|}{2.0} &1.0&1.0&1.0 &1.0 \\
		$e^+$ tracking                  &\multicolumn{2}{c|}{1.0} &\multicolumn{2}{c|}{1.0}&1.0 &1.0 &1.0 &1.0 \\
		$e^+$ PID                       &\multicolumn{2}{c|}{1.0} &\multicolumn{2}{c|}{1.0}&1.0 &1.0 &1.0 &1.0 \\
		Kinematic fit                     &\multicolumn{2}{c|}{1.7} &\multicolumn{2}{c|}{1.7}&1.7 &1.7 &1.7 &1.7 \\
		$E_{\rm extra \gamma}^{\rm max}$ and $N_{\rm char}^{\rm extra}$                   &\multicolumn{2}{c|}{0.7}  &\multicolumn{2}{c|}{0.7} &0.7  &0.7  &0.7 &0.7  \\
		Simultaneous fit to $M^2_{\rm miss}$                               &\multicolumn{2}{c|}{1.8} &\multicolumn{2}{c|}{1.5}&2.3 &2.5 &4.5 &2.2 \\
			\cline{2-3}
		$\pi^\pm/K^\pm$ tracking                &...  &2.0 &\multicolumn{2}{c|}{2.0}&2.0 &2.0 &...  &2.0 \\
		$\pi^\pm/K^\pm$ PID                     &...  &2.0 &\multicolumn{2}{c|}{2.0}&2.0 &2.0 &...  &2.0 \\
		\cline{4-5}
		$K^0_S$ reconstruction            &... &...  &...  &...  &...  &... &1.5 &...  \\
		MC  statistics                    &0.2 &0.4 &0.3 &0.3&0.3 &0.2 &0.2 &0.2 \\
		Quoted branching fractions                        &0.5 &1.2 &1.3 &1.4&1.1 &... &0.1 &...  \\
		MC model                          &0.7 &1.3 &1.2 &1.1&0.8 &2.4 &1.6 &0.9 \\
		Tag bias                          &0.8 &0.2 &0.5 &0.2&0.8 &0.7 &0.8 &0.5 \\
		Mass window                       &... &0.1 &0.1 &1.0&0.2 &0.2 &-- &1.2 \\
		\hline
		Total                             &\multicolumn{2}{c|}{4.3} &\multicolumn{2}{c|}{5.0}&5.1 &5.6 &6.0 &5.3\\
		\hline
		\hline
	\end{tabular}
\end{table*}

\section{Hadronic form factor}
To study the decay dynamics of $D^+_s \to he^+\nu_{e}$~($h=\eta$, $\eta^\prime$, or $K^0$),  the  candidate events for each signal decay are  divided into $N$~($N=$5 or 3) $q^2$ intervals. A least-$\chi^2$ fit
is performed  to the experimentally  measured ($\Delta\Gamma^i_{\rm msr}$) and theoretically expected ($\Delta\Gamma^i_{\rm th}$) differential decay rates  in the $i^{\rm th}$ $q^2$ interval~\cite{R:2015}. The $\Delta\Gamma^i_{\rm msr}$ in each interval are  determined as
$\Delta \Gamma^i_{\rm msr}= \frac{N_{\rm prd}^i}{\tau_{D_s^+}
	\cdot N_{\rm ST}}$, where
$\tau_{D_s^+}$ is the  lifetime of  $D_s^+$ ~\cite{pdg2024,Aaij:2017vqj}.
The number of events produced in data is calculated as
\begin{equation}\label{eq3}
	N_{\mathrm{prd}}^{i}=\sum_{j}^{N_{\mathrm{intervals}}}(\varepsilon\cdot {\mathcal B}_{\text{sub}})^{-1}_{ij}N_{\mathrm{DT}}^{j},
\end{equation}

where $N_{\rm DT}^j$ is the  signal yield observed in the $j$-th $q^{2}$ interval,  ${\mathcal B}_{\text{sub}}$ is the
product of the branching fractions of the intermediate decays in the signal decay,
and $\varepsilon$ is the efficiency matrix, which also includes the effects of bin migration,
given by
\begin{equation} \label{eq4}
	\varepsilon_{ij}=\sum_k
	\left[(N^{ij}_{\mathrm{rec}} \cdot N_{\rm ST})/(N^{j}_{\mathrm{gen}} \cdot \varepsilon_{\mathrm{ST}})\right]_k/N_{\rm ST}.
\end{equation}
Here, $N^{ij}_{\mathrm{rec}}$ is the  signal yield
generated in the $j$-th $q^{2}$ interval and reconstructed in the $i$-th $q^{2}$ interval,
$N^{j}_{\mathrm{gen}}$ is the total signal yield generated in the $j$-th $q^{2}$ interval, and the index $k$ sums over all tag modes and energies.
The signal yield $N^j_{\rm DT}$ in each $q^2$ interval is obtained from the fit to the corresponding
$M^2_{\mathrm{miss}}$ distribution.
The efficiency matrices   are shown in Tables~\ref{table:effmatrixetaev}, ~\ref{table:effmatrixetapev}, and ~\ref{table:effmatrixksoev}. Detailed information about the  $q^2$ divisions, as well as the  obtained   $N^i_{\rm DT}$, $N^i_{\rm prd}$, and $\Delta\Gamma^i_{\rm msr}$ of different $q^2$ intervals  for $D^+_s \to he^+\nu_{e}$ are shown in Tables~\ref{tab:decayrateb1}, ~\ref{tab:decayrateb2}, and ~\ref{tab:decayrateb3}.~\\

Using the values of $\Delta\Gamma^i_{\rm msr}$ obtained above and the theoretical parameterization of the partial decay rate $\Delta\Gamma^i_{\rm exp}$ described below, the parameters $r_1$ and $f^h_+(0)|V_{cq}|$  are obtained by minimizing the  $\chi^{2}$  constructed as
\begin{eqnarray}
	\label{eq:chisq}
	\chi^{2}=\sum_{i,j=1}^{N_{\mathrm{intervals}}}&(&\Delta\Gamma^{i}_{\mathrm{msr}}-\Delta\Gamma^{i}_{\mathrm{exp}})
	C_{ij}^{-1} \nonumber \\
	&(&\Delta\Gamma^{j}_{\mathrm{msr}}-\Delta\Gamma^{j}_{\mathrm{exp}}),
\end{eqnarray}
\hspace{-0.2cm} where $C_{ij} =
C_{ij}^{\mathrm{stat}}+C_{ij}^{\mathrm{syst}}$ is the covariance
matrix of the measured partial decay rates among $q^2$ intervals.

\begin{table*}[htbp]
	\centering
	\caption{The efficiency matrices for $\etaev$ averaged over all 14 ST modes, where $\varepsilon_{ij}$ represents the efficiency in \%
		for events produced in the $j$-th $q^2$ interval and reconstructed in the $i$-th $q^2$ interval. The efficiencies do not include the branching fractions of
		the $\eta$ decays  ($\mathcal B_{\rm sub}$), which are   (39.36$\pm$0.18)\% and (32.18$\pm$0.07)\% for $D^+_s \to \eta_{\gamma\gamma} e^+\nu_{e}$ and  $D^+_s \to \eta_{\pi^+\pi^-\pi^0} e^+\nu_{e}$ ~\cite{pdg2024}, respectively.
		 }
	\label{table:effmatrixetaev}
	\begin{tabular}{c|ccccc|ccccc}
		\hline\hline
	\multirow{2}{*}{$\varepsilon_{ij}$}	&	\multicolumn{5}{c|}{$D^+_s \to \eta_{\gamma\gamma} e^+\nu_{e}$}&	\multicolumn{5}{c}{$D^+_s \to \eta_{\pi^+\pi^-\pi^0} e^+\nu_{e}$}\\
		 &1&2&3&4&5&1&2&3&4&5\\
		\hline
		1 &47.57  &6.39 &2.35 &2.17 &2.08 &19.72  &1.81 &0.05 &0.00 &0.04\\
		2 &4.27 &38.67  &5.28 &0.33 &0.06 &1.84 &16.48  &2.34 &0.14 &0.10\\
		3 &0.37 &5.07 &35.60  &7.30 &0.93 &0.11 &2.32 &14.38  &3.10 &0.35\\
		4 &0.08 &0.46 &4.73 &31.15  &7.18 &0.02 &0.17 &2.07 &11.59  &2.67\\
		5 &0.12 &0.29 &1.30 &7.62 &38.26  &0.03 &0.08 &0.42 &3.20 &14.01\\
		\hline\hline
	\end{tabular}
\end{table*}

\begin{table}[htbp]
	\centering
	\caption{The efficiency matrices for $\etapev$ averaged over all 14 ST modes, where $\varepsilon_{ij}$ represents the efficiency in \%
		for events produced in the $j$-th $q^2$ interval and reconstructed in the $i$-th $q^2$ interval. The efficiencies do not include the branching fractions of
		the $\eta^\prime$ decays ($\mathcal B_{\rm sub}$), which are   (16.72$\pm$0.30)\% and (29.5$\pm$0.4)\% for $D^+_s \to \eta^{\prime}_{\pi^+\pi^-\eta} e^+\nu_{e}$ and $D^+_s \to \eta^\prime_{\pi^+\pi^-\gamma} e^+\nu_{e}$~\cite{pdg2024}, respectively.
		 }
	\label{table:effmatrixetapev}
	\begin{tabular}{c|ccc|ccc}
		\hline\hline
		\multirow{2}{*}{$\varepsilon_{ij}$}	&	\multicolumn{3}{c|}{$D^+_s \to \eta^{\prime}_{\pi^+\pi^-\eta} e^+\nu_{e}$}&	\multicolumn{3}{c}{$D^+_s \to \eta^\prime_{\pi^+\pi^-\gamma} e^+\nu_{e}$}\\
		&1&2&3&1&2&3\\
		\hline
		1 &20.17  &2.39 &0.11 &28.42  &3.47 &0.21 \\
		2 &2.30 &16.52  &2.51 &3.46 &24.19  &3.66\\
		3 &0.30 &3.38 &18.96  &0.49 &4.92 &28.66 \\
		\hline\hline
	\end{tabular}
\end{table}

\begin{table}[htbp]
	\centering
	\caption{The efficiency matrix for $\Koev$ averaged over all 14 ST modes, where $\varepsilon_{ij}$ represents the efficiency in \%
		for events produced in the $j$-th $q^2$ interval and reconstructed in the $i$-th $q^2$ interval. The efficiencies do not include the  branching fraction of 
		the $K^0$ decay  ($\mathcal B_{\rm sub}$), which is $(34.60\pm0.03)\%$~\cite{pdg2024}.
		}
	\label{table:effmatrixksoev}
	\begin{tabular}{c|ccc}
		\hline\hline
		$\varepsilon_{ij}$&1&2&3\\
		\hline
		1 &43.33  &3.99 &0.06 \\
		2 &3.53 &38.74  &3.19 \\
		3 &0.17 &4.89 &40.92  \\
		\hline\hline
			\end{tabular}
\end{table}

\begin{table*}[htbp]\centering
	\caption{
		The partial decay rates of  $\etaev$ in different $q^{2}$ intervals, where the uncertainties are statistical only. }
	\label{tab:decayrateb1}
		\begin{tabular}{ccccccc}\hline\hline
			$i$&1&2&3&4&5&\multirow{2}{*}{Sum}\\
			$q^2$ $(\mathrm{GeV}^{2}/c^{4})$&($0,\,0.40$)&($0.40,\,0.80$)&($0.80,\,1.20$)&($1.20,\,1.50$)&($1.50,\,2.02$)&\\
			\hline
			$N_{\mathrm{DT}}^i$&239.9$\pm$19.7&212.7$\pm$18.8&144.5$\pm$14.1&76.0$\pm$8.7&48.5$\pm$9.4&721.6$\pm$33.3\\
			$N_{\mathrm{prd}}^i$&872$\pm$87&937$\pm$104&612$\pm$88&361$\pm$66&163$\pm$55&2945$\pm$174\\
			$\Delta\Gamma^i_{\rm msr}$ $(\mathrm{ns^{-1}})$&14.0$\pm$1.4&15.1$\pm$1.7&9.8$\pm$1.4&5.8$\pm$1.1&2.6$\pm$0.9&47.3$\pm$2.2\\
			\hline
			
			\hline\hline
		\end{tabular}
\end{table*}

\begin{table*}[htbp]\centering
	\caption{
		The partial decay rates of  $\etapev$ in different $q^{2}$ intervals, where the uncertainties are statistical only. }
	\label{tab:decayrateb2}
	\begin{tabular}{ccccc}\hline\hline
	
		$i$&1&2&3&\multirow{2}{*}{Sum}\\
		$q^2$ $(\mathrm{GeV}^{2}/c^{4})$&($0,\,0.25$)&($0.25,\,0.50$)&($0.50,\,1.02$)&\\
		\hline
		$N_{\mathrm{DT}}^i$&57.0$\pm$9.8&50.5$\pm$9.7&30.9$\pm$7.6&138.4$\pm$15.8\\
		$N_{\mathrm{prd}}^i$&433$\pm$86&420$\pm$103&186$\pm$69&1039$\pm$151\\
		$\Delta\Gamma^i_{\rm msr}$ $(\mathrm{ns^{-1}})$&7.0$\pm$1.4&6.8$\pm$1.7&3.0$\pm$1.1&16.8$\pm$2.0\\
	
		\hline\hline
	\end{tabular}
\end{table*}

\begin{table*}[htbp]\centering
	\caption{
		The partial decay rates of  $\Koev$ in different $q^{2}$ intervals, where the uncertainties are statistical only. }
	\label{tab:decayrateb3}
	\begin{tabular}{ccccc}\hline\hline
		$i$&1&2&3&\multirow{2}{*}{Sum}\\
		$q^2$ $(\mathrm{GeV}^{2}/c^{4})$&($0,\,0.45$)&($0.45,\,0.90$)&($0.90,\,2.16$)&\\
		\hline
		$N_{\mathrm{DT}}^i$&20.3$\pm$5.0&14.9$\pm$4.9&16.0$\pm$4.0&51.2$\pm$8.1\\
		$N_{\mathrm{prd}}^i$&127$\pm$34&91$\pm$38&101$\pm$25&319$\pm$57\\
		$\Delta\Gamma^i_{\rm msr}$ $(\mathrm{ns^{-1}})$&2.0$\pm$0.5&1.5$\pm$0.6&1.6$\pm$0.4&5.1$\pm$0.9\\		
		\hline\hline
	\end{tabular}
\end{table*}

\begin{figure*}[htbp]\centering
	\hspace{-1.2cm}
	\includegraphics[width=0.8\linewidth]{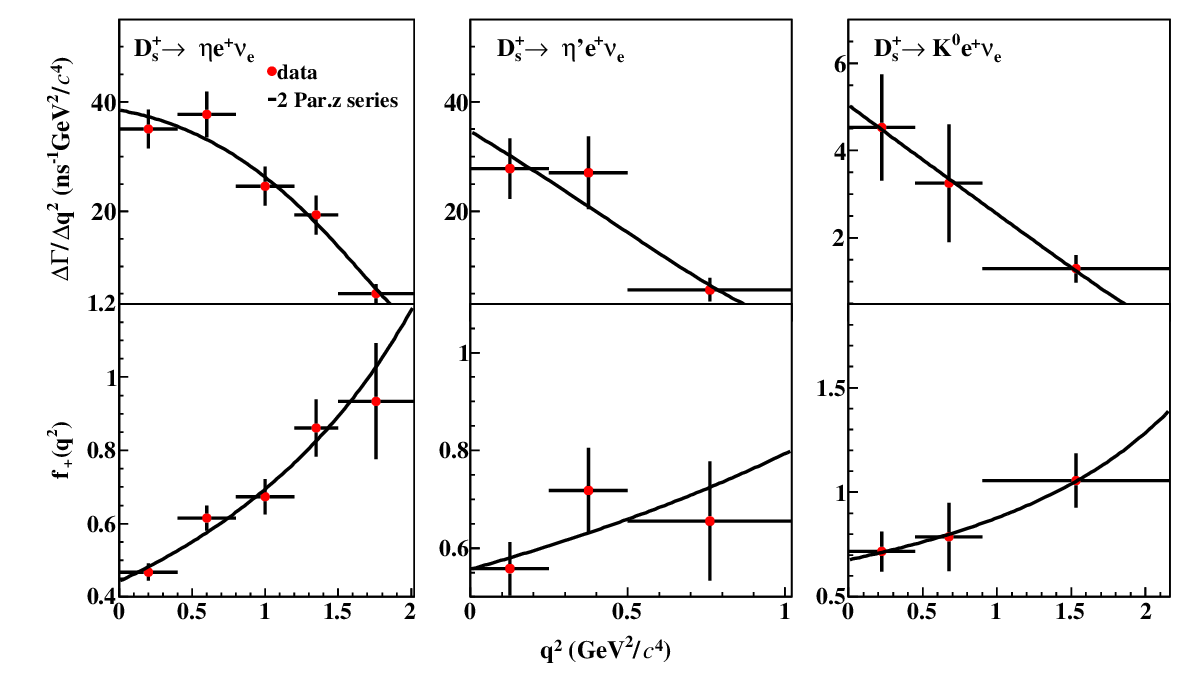}
	\caption{
		(Top) Fits to the partial decay rates of
		the  semileptonic decays  $\etaev$, $\etapev$, and $\Koev$ and (bottom)
		projections on the hadronic form factor as a function of $q^2$.
		The dots with error bars are the measured partial decay rates
		and the solid curves are the fits.}
	\label{fig:fitdecayrate}
\end{figure*}

For each signal decay, its  differential decay rate can be written as~\cite{kang}
\begin{equation}
	\frac{d\Gamma(D_s^+\to he^+\nu_e)}{dq^2}=\frac{G^2_F|V_{cs(d)}|^2}{24\pi^3}p^3_{h}|f^{h}_+(q^2)|^2,
	\label{eq:dgammadq2_ksev}
\end{equation}
where  $G_F$ is the Fermi coupling constant~\cite{pdg2024}, $p_{h}$ is the momentum of $h$ in the $D_s^+$ rest frame and the  positron mass is neglected. The hadronic FF $f_+^{h}(q^2)$ is described by using  the two-parameter series expansion model, which can be written as
\begin{equation}
	f^{h}_{+}(q^2)=\frac{f^{h}_{+}(0)P(0)\Phi(0,t_{0})}{P(q^2)\Phi(q^2,t_{0})}\cdot \frac{1+r_{1}(t_{0})z(q^2,t_{0})}{1+r_{1}(t_{0})z(0,t_{0})},
\end{equation}
where
$t_{0}=t_{+}(1-\sqrt{1-t_{-}/t_{+}})$, $t_{\pm}=(m_{D^+_s}\pm m_{h})^{2}$, and
the functions $P(q^2)$, $\Phi(q^2, t_0)$, and $z(q^2, t_0)$ are defined following Ref.~\cite{Becher:2005bg}.

For $D^+_s\to \eta e^+\nu_e$ and $D^+_s\to \eta^\prime e^+\nu_e$, the two reconstructed modes of $\eta$ or $\eta^\prime$ have been combined in the determining partial decay rates, where the signal efficiencies have been re-weighted by individual branching fractions. We construct the  statistical and systematic covariance matrices to be
$C_{ij}^{\rm stat} = (\frac{1}{\tau_{D_s^+}N_{\rm ST}})^2\sum_{\alpha}\epsilon_{i\alpha}^{-1}\epsilon_{j\alpha}^{-1}[\sigma(N^{\alpha}_{\rm
	DT})]^2$ and
$C_{ij}^{\rm syst} = \delta(\Delta \Gamma^i_{\rm msr})\delta(\Delta \Gamma^j_{\rm msr})$,
respectively,
where $\sigma(N^\alpha_{\rm DT})$ and $\delta(\Delta\Gamma^i_{\rm msr})$
are the statistical and systematic uncertainties in the $\alpha_{\rm th}$ and  $i_{\rm th}$ $q^2$ intervals, respectively.
The sources of the systematic uncertainties are almost the same as branching fraction measurement, except for  an additional systematic uncertainty of 0.4\% from the $D^+_s$ lifetime,   $\tau_{D_s^+}$~\cite{pdg2024}, is included. The systematic uncertainty due to form factor parameterization  is assigned as the difference of the fitted results for $D^+_s\to \eta e^+\nu_e$ between the fits with two-parameter or three-parameter parameterizations. The same systematic uncertainty is assigned for  $D^+_s\to \eta^\prime e^+\nu_e$ and $D^+_s\to K^0e^+\nu_e$   due to limited statistics. The $C_{ij}^{\rm syst}$ is obtained by summing the covariance matrices
for all systematic uncertainties. statistical and systematic covariance density matrices for $D^+_s\to \eta e^+\nu_e$, 
$D^+_s\to \eta^\prime e^+\nu_e$, and $D^+_s\to K^0 e^+\nu_e$ are summarized in Tables~\ref{table:statmatrixa},~\ref{table:statmatrixb}, and ~\ref{table:statmatrixc}, respectively.

\begin{table*}[htbp]\centering
	\caption{Statistical and systematic covariance density matrices for  $\etaev$ in different
			$q^2$ intervals.}
	\label{table:statmatrixa}
	\begin{tabular}{cccccc|cccccc}\hline\hline
			\multicolumn{6}{c|}{Statistical part}&\multicolumn{6}{c}{Systematic part}\\		
		$\rho^{\rm stat}_{ij}$	&1      &2      &3      &4      &5  	& $\rho^{\rm syst}_{ij}$	&1      &2      &3 	&4      &5    \\
			\hline
			1 &1.000  &-0.230 &0.016  &-0.014 &-0.015 & 1 &1.000  &0.857  &0.926  &0.939  &0.967  \\
			2 &-0.230 &1.000  &-0.280 &0.051  &-0.010 &2 &0.857  &1.000  &0.700  &0.935  &0.886  \\
			3 &0.016  &-0.280 &1.000  &-0.343 &0.047 &3 &0.926  &0.700  &1.000  &0.861  &0.915  \\
			4 &-0.014 &0.051  &-0.343 &1.000  &-0.403&4 &0.939  &0.935  &0.861  &1.000  &0.951  \\
			5 &-0.015 &-0.010 &0.047  &-0.403 &1.000 &5 &0.967  &0.886  &0.915  &0.951  &1.000  \\
			
			\hline\hline
		\end{tabular}
\end{table*}

\begin{table*}[htbp]\centering
	\caption{Statistical and systematic covariance density matrices for  $\etapev$ in different
		$q^2$ intervals.}
	\label{table:statmatrixb}
	\begin{tabular}{cccc|cccc}\hline\hline
		\multicolumn{4}{c|}{Statistical part }&\multicolumn{4}{c}{Systematic  part}\\		
		$\rho^{\rm stat}_{ij}$	&1      &2      &3   & $\rho^{\rm syst}_{ij}$     	&1      &2      &3    \\
		\hline
	1 &1.000  &-0.258 &0.055  & 1 &1.000  &0.925  &0.712\\
	2 &-0.258 &1.000  &-0.347 & 2 &0.925  &1.000  &0.717  \\
	3 &0.055  &-0.347 &1.000  & 3 &0.712  &0.717  &1.000 \\
		\hline\hline
	\end{tabular}
\end{table*}

\begin{table*}[htbp]\centering
	\caption{Statistical and systematic covariance density matrices for  $\Koev$ in different
		$q^2$ intervals.}
	\label{table:statmatrixc}
	\begin{tabular}{cccc|cccc}\hline\hline
		\multicolumn{4}{c|}{Statistical part}&\multicolumn{4}{c}{Systematic part}\\		
		$\rho^{\rm stat}_{ij}$	&1      &2      &3    & $\rho^{\rm syst}_{ij}$     	&1      &2      &3    \\
		\hline
	1 &1.000  &-0.183 &0.032& 1 &1.000  &0.905  &0.962   \\
	2 &-0.183 &1.000  &-0.233& 2 &0.905  &1.000  &0.868   \\
	3 &0.032  &-0.233 &1.000 & 3 &0.962  &0.868  &1.000  \\
	
		\hline\hline
	\end{tabular}
\end{table*}

For each decay, the fit to their corresponding partial decay rates in different $q^2$ intervals gives the fitted parameters $f^h_+(0)|V_{cq}|$ and $r_1$. 
The final  fit
results are shown in Fig.~\ref{fig:fitdecayrate} and the obtained parameters  are summarized in Table~\ref{tab:sum_formfactor}. The 
 nominal fit parameters
are taken from the fit with the combined statistical and systematic
covariance matrix, and their statistical uncertainties are taken from
the fit with the statistical covariance matrix. For each
parameter, the systematic uncertainty is obtained by calculating the
quadratic difference of uncertainties between these two fits.
Taking the CKM matrix element $|V_{cs}| = 0.97320\pm0.00011$ and $|V_{cd}| = 0.22486\pm0.00067$~\cite{pdg2024} as input,
we determine  $f^{h}_+(0)$ for each signal decay. The obtained results are  summarized in the last column of Table~\ref{tab:sum_formfactor}, where the first uncertainties are statistical and the second are systematic.

\begin{table*}
	\centering
	\caption{The obtained parameters of the hadronic form factors for  $\etaev$, $\etapev$, and $\Koev$. The first uncertainties are statistical and the second systematic. 
		The $\rho_{f^h_+(0)|V_{cq}|}$ is the correlation coefficient between $r_1$
		and $f^{h}_+(0)|V_{cq}|$.
		The NDF denotes the number of degrees of
		freedom.
	}
	\label{tab:sum_formfactor}
		\begin{tabular}{c|c|c|c|c|c}
			\hline\hline
			Signal decay&$f^{h}_+(0)|V_{cq}|$& $r_1$&$\rho_{f^h_+(0)|V_{cq}|}$&$\chi^2/\rm NDF$&$f^{h}_+(0)$\\
			\hline 
			$\etaev$ &$\fvzetaevnewnew$& $\rzetaevnewnew$ & 0.72&$\czetaevnew$ &$\fzetaevnewnew$\\
			$\etapev$&$\fvzetapevnewnew$& $\rzetapevnewnew$&0.82& $\czetapevnew$ &$\fzetapevnewnew$\\
			$\Koev$  &$\fvzkoevnewnew$ &  $\rzkoevnewnew$&0.83& $\czkoevnew$ &$\fzkoevnewnew$\\
			\hline
			\hline
		\end{tabular}
	
\end{table*}

\section{Summary}

\begin{figure*}[!htbp]
	\includegraphics[width=0.45\linewidth]{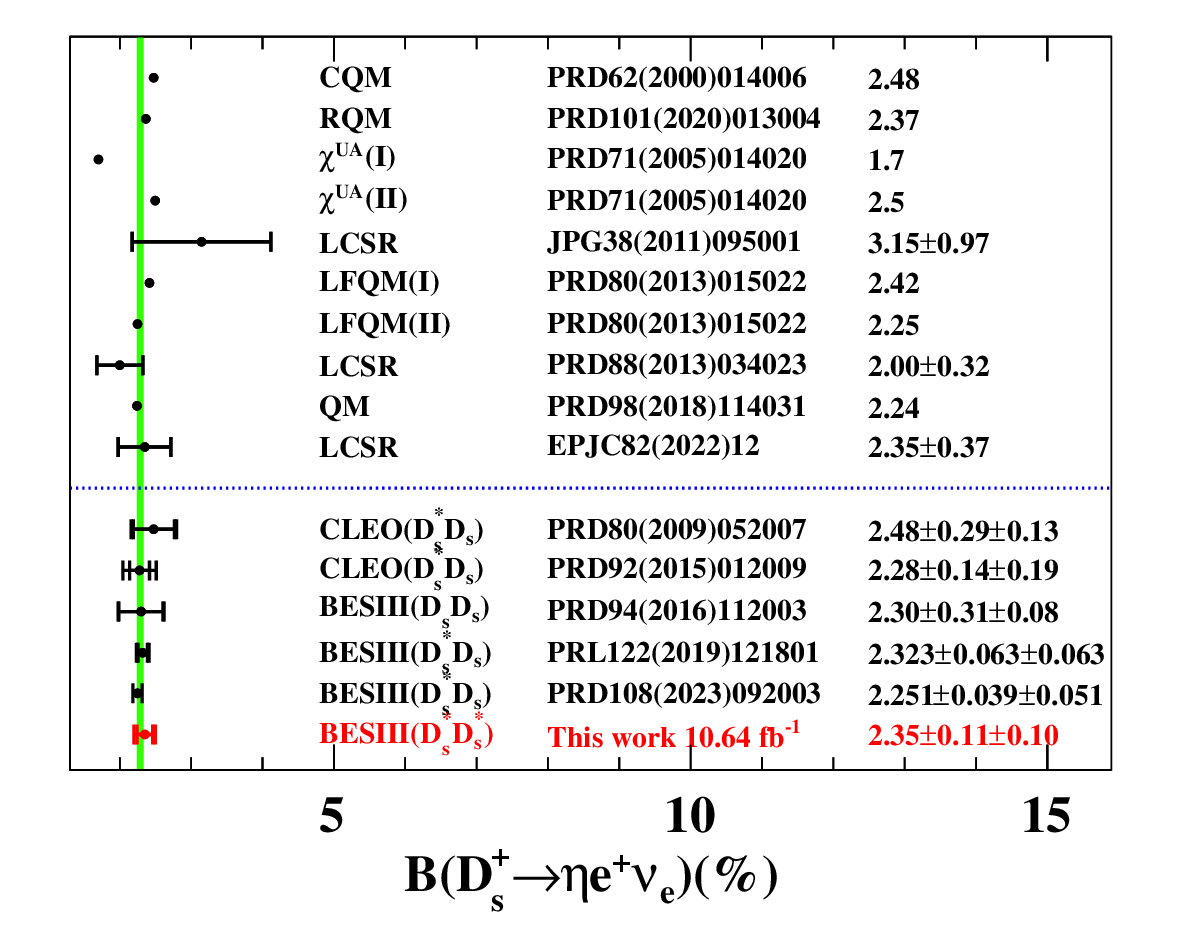}
	\includegraphics[width=0.45\linewidth]{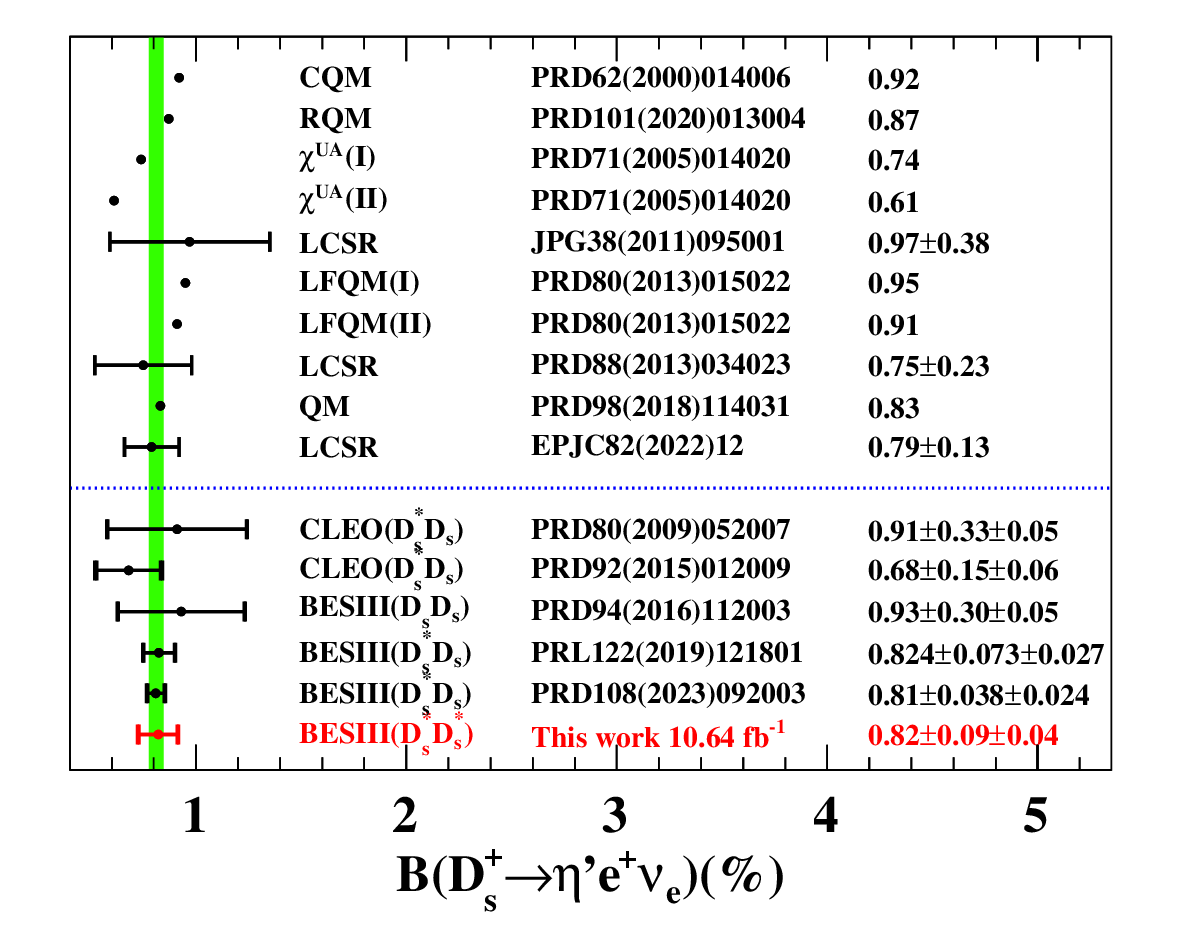}

	\includegraphics[width=0.45\linewidth]{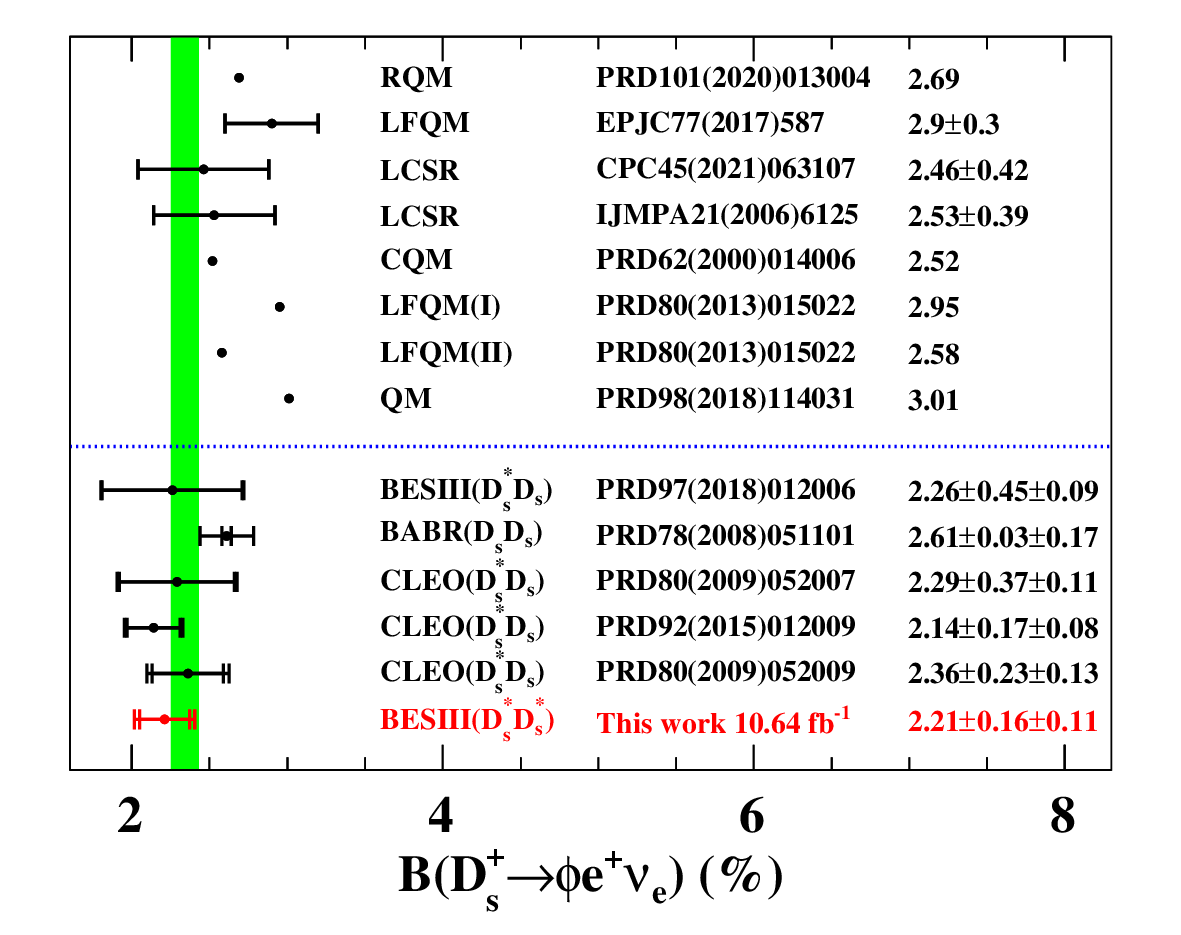}
	\includegraphics[width=0.45\linewidth]{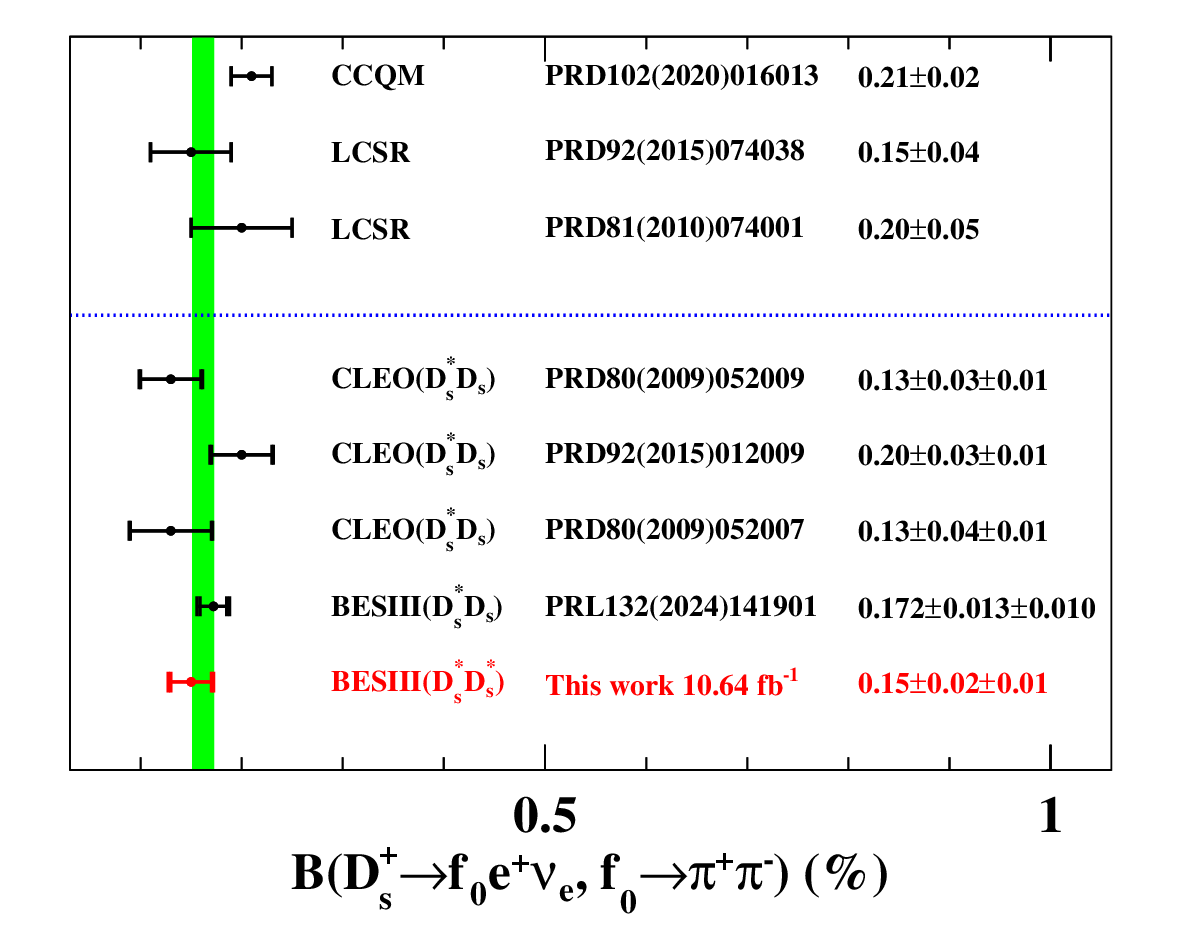}

	\includegraphics[width=0.45\linewidth]{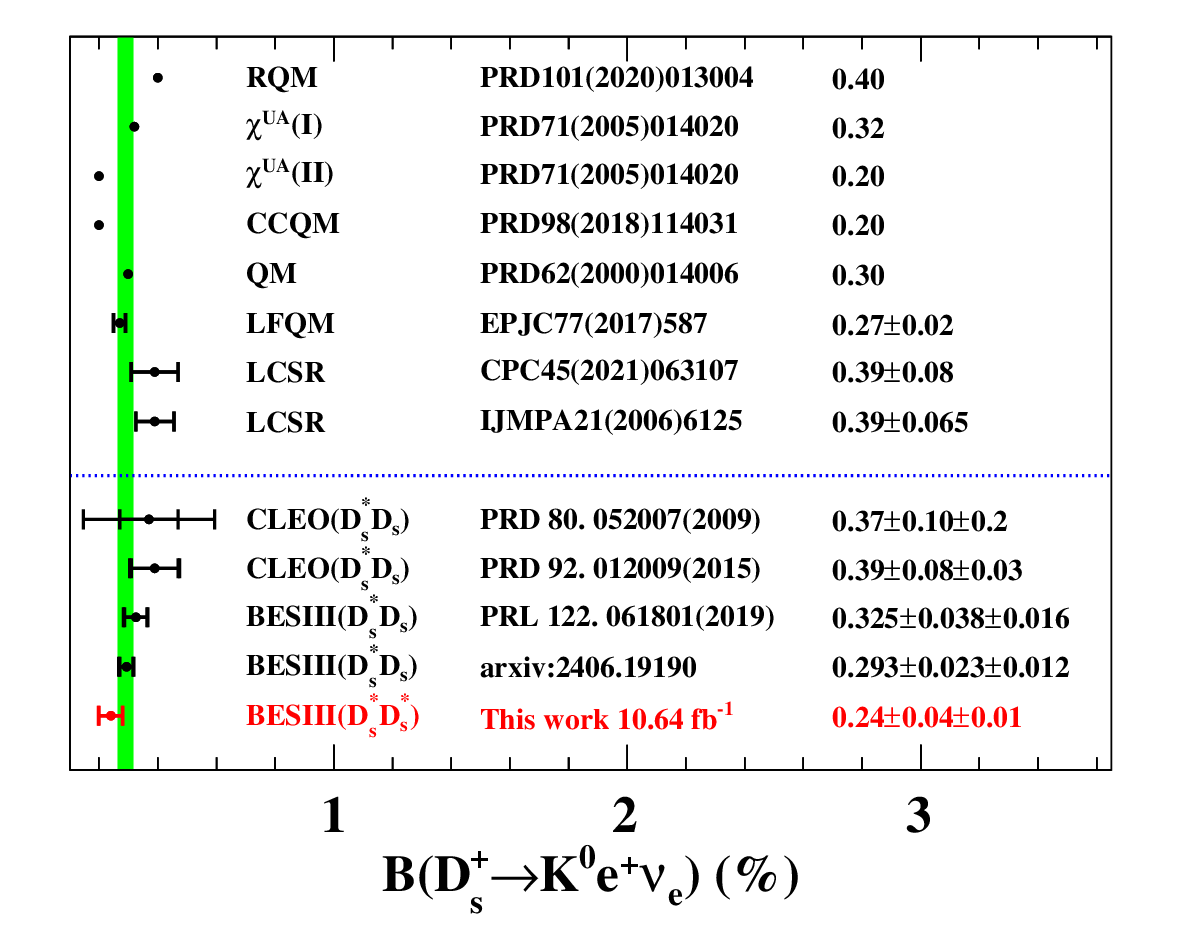}
	\includegraphics[width=0.45\linewidth]{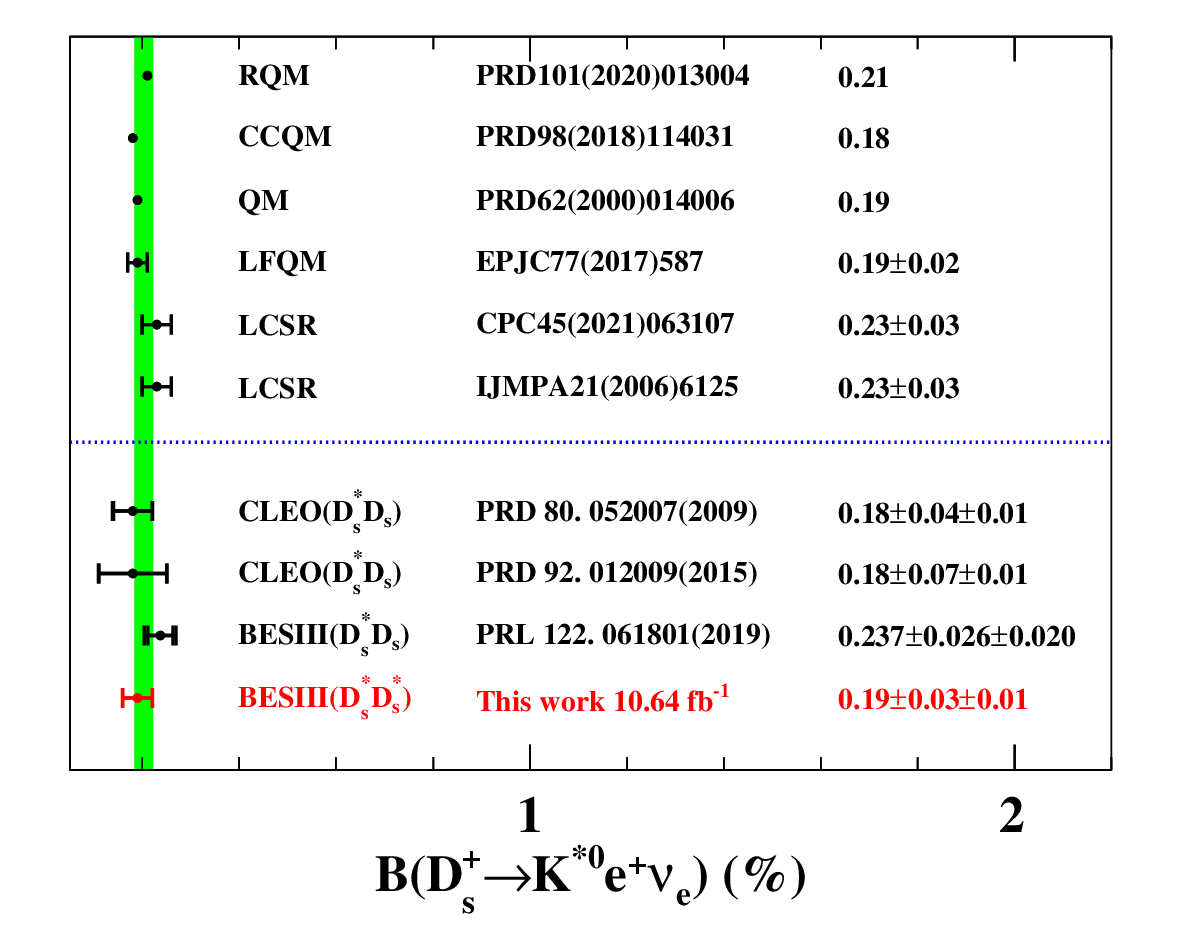}
\caption{
Comparisons of the branching fractions of semileptonic $D^+_s$ decays
with theoretical calculations and previous  experimental measurements. 
The $D_sD_s$, $D_s^*D_s$, and $D_s^*D_s^*$ in the brackets denote the measurements are made based on $e^+e^-\to D_s^+D_s^-$, $D_s^{*\pm}D_s^\mp$, and $D_s^{*+}D_s^{*-}$, respectively. The green bands correspond to the $\pm1\sigma$ limit of the world average include the results of this work. 
\label{fig:compare_Ds_semi}}
\end{figure*}

\begin{figure*}[!htbp]
	\includegraphics[width=0.45\linewidth]{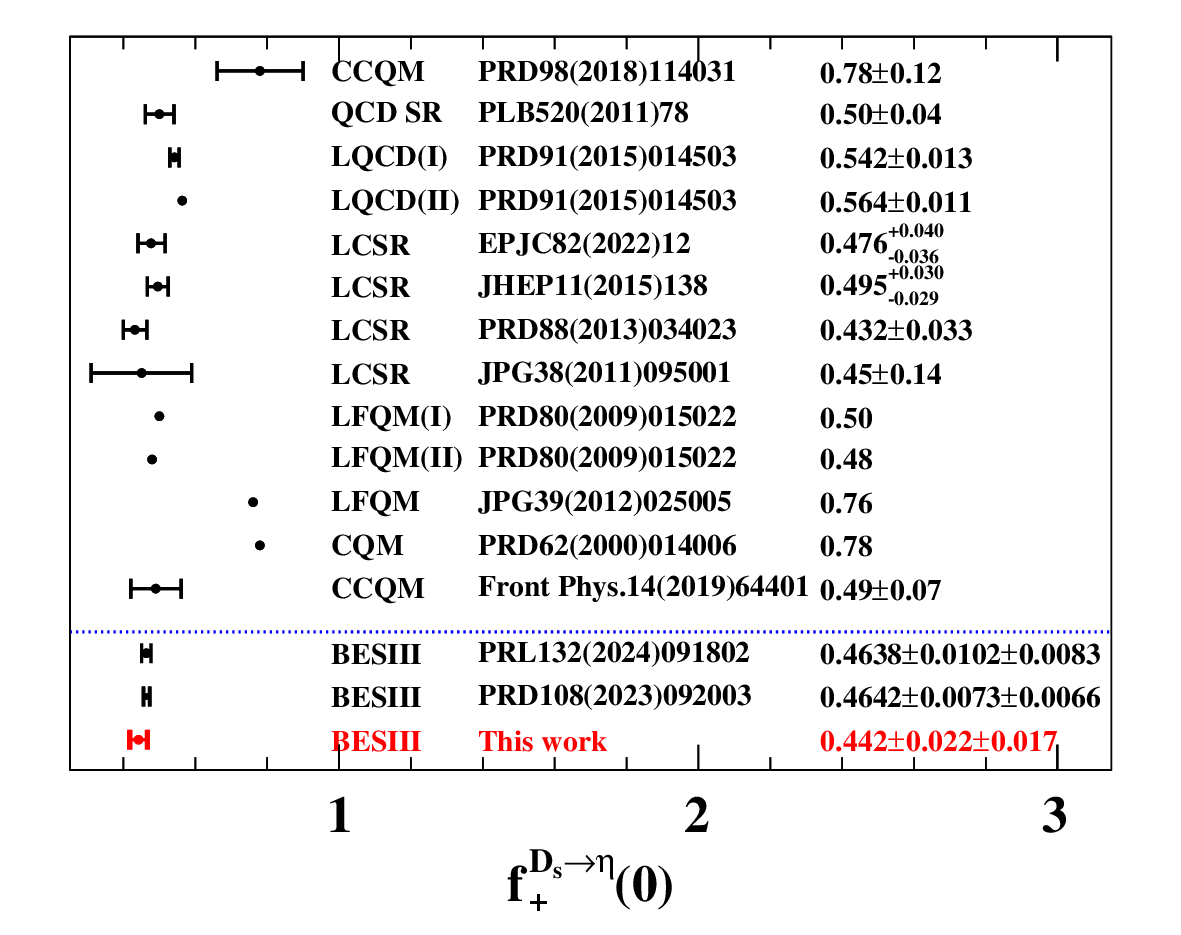}
	\includegraphics[width=0.45\linewidth]{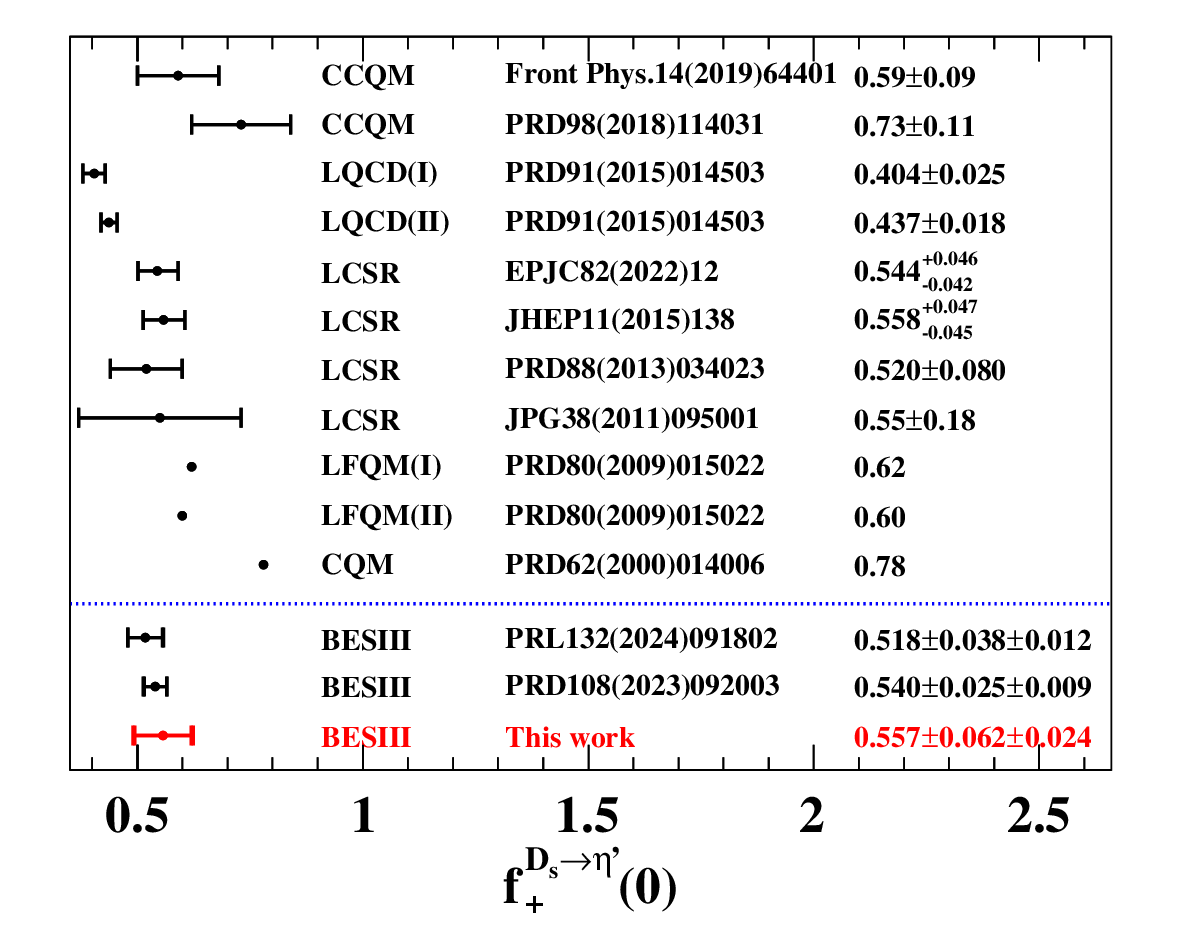}
	\includegraphics[width=0.45\linewidth]{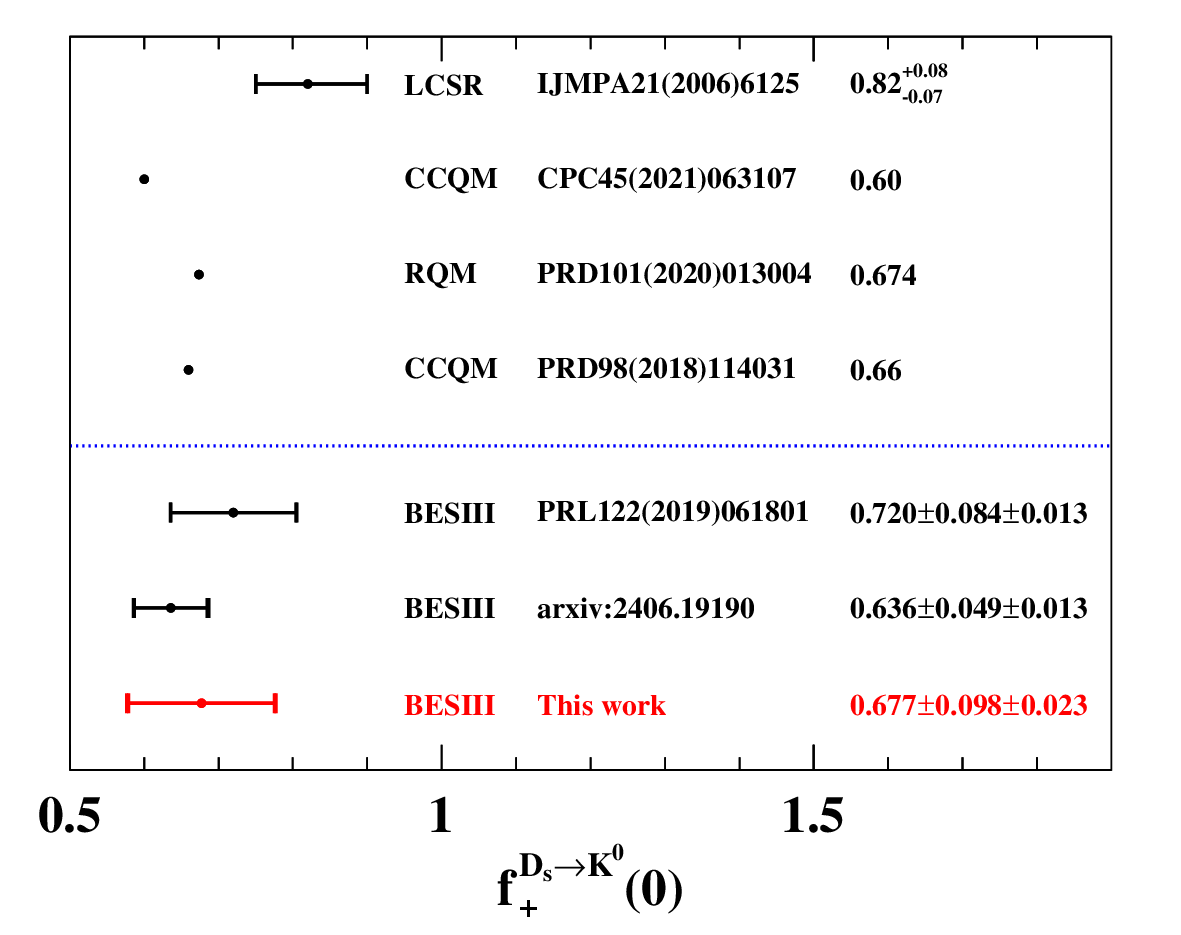}
	\caption{
		Comparisons of the form factors $f^{\eta}_+(0)$, $f^{\eta^\prime}_+(0)$, and  $f^{K^0}_+(0)$ measured by this work with the theoretical calculations  and previous  experimental measurements. The first and second uncertainties are statistical and systematic, respectively.
		\label{fig:compare_Ds_semi_factor}}
\end{figure*}

Using $10.64~\mathrm{fb}^{-1}$ of $e^+e^-$ collision data collected
with the BESIII detector at center-of-mass energies between 4.237 and
4.699 GeV, we report the  measurements of the branching fractions
of semileptonic $D^+_s$ decays via the $e^+e^-\to D_s^{*+}D_s^{*-}$
process. The obtained branching fractions are 
${\mathcal B}(D_s^+\to \eta e^+\nu_e)=(2.35\pm0.11_{\rm stat}\pm0.10_{\rm syst})\%,$ ${\mathcal
	B}(D_s^+\to \eta^\prime e^+\nu_e)=(0.82\pm0.09_{\rm stat}\pm0.04_{\rm syst})\%,$ ${\mathcal B}(D_s^+\to \phi e^+\nu_e)=(2.21\pm0.16_{\rm stat}\pm0.11_{\rm syst})\%,$
${\mathcal B}(D_s^+\to f_0(980)
e^+\nu_e,f_0(980)\to\pi^+\pi^-)=(0.15\pm0.02_{\rm stat}\pm0.01_{\rm syst})\%,$ ${\mathcal
	B}(D_s^+\to K^0 e^+\nu_e)=(0.24\pm0.04_{\rm stat}\pm0.01_{\rm syst})\%,$ and ${\mathcal B}(D_s^+\to K^{*0} e^+\nu_e)=(0.19\pm0.03_{\rm stat}\pm0.01_{\rm syst})\%.$ Figure~\ref{fig:compare_Ds_semi} shows comparisons of the
branching fractions of different signal decays with the theoretical calculations  and previous  experimental measurements.  The precisions of the branching fractions measured in
this work are not better than those measured via $e^+e^-\to
D_s^{*\pm}D_s^{\mp}$ with 7.33 fb$^{-1}$ of $e^+e^-$ collision data
taken between 4.128 and 4.226 GeV at BESIII. However, the precisions
are better than those measured via $e^+e^-\to
D_s^{*\pm}D_s^{\mp}$ with 0.6 fb$^{-1}$ of $e^+e^-$ collision data
taken at 4.17~GeV.  Using the two-parameter series expansion,
the hadronic  form factors of  $D^+_s\to \eta e^+\nu_e$, $D^+_s\to \eta^\prime e^+\nu_e$, and $D^+_s\to K^0 e^+\nu_e$ at $q^2=0$ are   determined to be $f^{\eta}_+(0) = 0.442\pm0.022_{\rm stat}\pm 0.017_{\rm syst},$
$f^{\eta^{\prime}}_+(0) = 0.557\pm 0.062_{\rm stat}\pm0.024_{\rm syst},$ and 
$f^{K^0}_+(0) = 0.677\pm0.098_{\rm stat}\pm0.023_{\rm syst}.$ Figure~\ref{fig:compare_Ds_semi_factor} shows comparisons of the
form factors of different signal decays with the theoretical calculations  and previous  experimental measurements. These  results  offer additional data to test different theoretical calculations on these hadronic form factors.

\section{Acknowledgment}

The BESIII Collaboration thanks the staff of BEPCII and the IHEP computing center for their strong support. This work is supported in part by National Key R\&D Program of China under Contracts Nos. 2023YFA1606000, 2023YFA1606704, 2020YFA0406300, 2020YFA0406400; National Natural Science Foundation of China (NSFC) under Contracts Nos. 12375092, 11635010, 11735014, 11935015, 11935016, 11935018, 11961141012, 12025502, 12035009, 12035013, 12061131003, 12192260, 12192261, 12192262, 12192263, 12192264, 12192265, 12221005, 12225509, 12235017; the Chinese Academy of Sciences (CAS) Large-Scale Scientific Facility Program; the CAS Center for Excellence in Particle Physics (CCEPP); Joint Large-Scale Scientific Facility Funds of the NSFC and CAS under Contract No. U1832207; 100 Talents Program of CAS; The Institute of Nuclear and Particle Physics (INPAC) and Shanghai Key Laboratory for Particle Physics and Cosmology; German Research Foundation DFG under Contracts Nos. 455635585, FOR5327, GRK 2149; Istituto Nazionale di Fisica Nucleare, Italy; Ministry of Development of Turkey under Contract No. DPT2006K-120470; National Research Foundation of Korea under Contract No. NRF-2022R1A2C1092335; National Science and Technology fund of Mongolia; National Science Research and Innovation Fund (NSRF) via the Program Management Unit for Human Resources \& Institutional Development, Research and Innovation of Thailand under Contract No. B16F640076; Polish National Science Centre under Contract No. 2019/35/O/ST2/02907; The Swedish Research Council; U. S. Department of Energy under Contract No. DE-FG02-05ER41374.

\end{document}